\pdfoutput=1
\documentclass[camera]{jpaper}
\usepackage[sort, compress]{cite}
\usepackage{soul} 
\usepackage{mathtools}
\usepackage{amsmath}
\usepackage{graphicx}
\usepackage[linesnumbered,ruled]{algorithm2e}

\usepackage{algpseudocode}
\usepackage{titlesec}
\usepackage{microtype} 

\algrenewcommand\algorithmiccomment[2][\normalsize]{{#1\hfill\(\triangleright\) #2}}
\usepackage{geometry} \usepackage{array}

\makeatletter
\let\MYcaption\@makecaption
\makeatother

\usepackage[font=footnotesize]{subcaption}

\makeatletter
\let\@makecaption\MYcaption
\makeatother

\usepackage{fixltx2e}
\usepackage{dblfloatfix}

\usepackage[usenames,dvipsnames,table]{xcolor}
\usepackage{blindtext}

\usepackage{xspace}
\usepackage{ifthen}

\usepackage{booktabs}
\usepackage[all]{nowidow}
\usepackage{multirow}
\usepackage{comment}
\usepackage{enumitem}
\usepackage{setspace}
\usepackage{booktabs}
\usepackage{multirow}
\usepackage{amssymb}
\usepackage{url}

\newcommand{\squishend}{
  \end{list}  }

\usepackage{tikz}
\usetikzlibrary{arrows}
\usetikzlibrary{patterns}
\usetikzlibrary{calc}
\usetikzlibrary{shapes}
\usetikzlibrary{positioning,fit}
\usepackage{float}
\usepackage{lipsum}

\usepackage[nolessnomore, italic]{mathastext}
\usepackage[T1]{fontenc}
\usepackage{hhline}
\usepackage[normalem]{ulem}
\usepackage{indentfirst}
\usepackage{footmisc}
\usepackage{datetime}

\newcommand{\incircle}[1]{\raisebox{0.5pt}{\protect%
  \tikz[baseline=(char.base)]{
  \node[shape=circle,draw,inner sep=0.5pt,minimum height=10pt, fill=black,text=white,font=\footnotesize\sffamily] (char) {#1};}%
}}%
\newcommand{\inhollowcircle}[1]{\raisebox{0.5pt}{\protect%
  \tikz[baseline=(char.base)]{
  \node[shape=circle,draw,inner sep=0.5pt,minimum height=10pt,font=\footnotesize\sffamily] (char) {#1};}%
}}%
\newcommand{\secref}[1]{\S\ref{#1}}

\newif\ifcameraready
\camerareadyfalse

\definecolor{amber}{rgb}{1.0, 0.49, 0.0}
\definecolor{darkgreen}{rgb}{0.0, 0.2, 0.13}
\definecolor{darkbyzantium}{rgb}{0.36, 0.22, 0.33}
\definecolor{darkseagreen}{rgb}{0.56, 0.74, 0.56}
\definecolor{darkspringgreen}{rgb}{0.09, 0.45, 0.27}
\definecolor{dollarbill}{rgb}{0.52, 0.73, 0.4}

\definecolor{darkcyan}{rgb}{0.0, 0.55, 0.55}
\definecolor{forestgreen}{rgb}{0.0, 0.27, 0.13}
\definecolor{azure}{rgb}{0.0, 0.5, 1.0}

\newcommand{\todo}[1]{}
\renewcommand{\todo}[1]{{\color{red} TODO: {#1}}}
\newcommand{\nas}[1]{\textcolor{black}{#1}}

\newcommand{\sgdel}[1]{}

\newcommand{\rebuttdel}[1]{}

\newcommand{\cameradelete}[1]{}

\newcommand{\shepherddel}[1]{}
\newcommand{\sgaddi}[1]{}

\newcommand{\sg}[1]{\textcolor{black}{#1}}

\newcommand{\onurii}[1]{\textcolor{black}{#1}}
\newcommand{\onurthird}[1]{\textcolor{black}{#1}}
\newcommand{\nasi}[1]{\textcolor{black}{#1}}
\newcommand{\sgii}[1]{\textcolor{black}{#1}}
\newcommand{\onurv}[1]{\textcolor{black}{#1}}

\definecolor{amber}{rgb}{1.0, 0.49, 0.0}

\newcommand{\mpt}[1]{{\color{black}#1}}
\newcommand{\mhp}[1]{{\color{black}#1}} 
\newcommand{\pointer}[1]{{\color{black}#1}}
\newcommand{\onurvi}[1]{\textcolor{black}{#1}}
\newcommand{\onurvii}[1]{\textcolor{black}{#1}}

\newcommand{\sysfull}{Virtual Block Interface\xspace}
\newcommand{\sys}{VBI\xspace}

\newcommand{\xeightsix}{x86-64\xspace}

\newcommand{\requestvas}{\texttt{request\_vb}\xspace}

\newcommand{\enablevb}{{\texttt{enable\_vb}}\xspace}
\newcommand{\disablevb}{{\texttt{disable\_vb}}\xspace}

\newcommand{\attachvb}{{\texttt{attach}}\xspace}
\newcommand{\detachvb}{{\texttt{detach}}\xspace}

\newcommand{\clonevb}{{\texttt{clone\_vb}}\xspace}
\newcommand{\promotevb}{{\texttt{promote\_vb}}\xspace}

\newcommand{\cvt}{CVT\xspace}

\newcommand{\vit}{VB Info Table\xspace}
\newcommand{\vitshort}{VIT\xspace}

\newcommand{\mtlfull}{Memory Translation Layer\xspace}
\newcommand{\mtl}{MTL\xspace}

\fancypagestyle{firstpage}{
  \fancyhf{}

  \fancyhead[C]{\textcolor{gray}{\large\emph{To Appear at ISCA 2020}}} 
  \fancyfoot[C]{\thepage}
}

\begin{document}
\newcommand{\affilSFU}{$^{\dagger}$}
\newcommand{\affilETH}{$^{\star}$}
\newcommand{\affilWA}{$^{\Join}$}
\newcommand{\affilCMU}{$^{\ddag}$}
\newcommand{\affilKING}{$^{\odot}$}
\newcommand{\affilBU}{$^{\diamond}$}
\newcommand{\affilMS}{$^{\triangledown}$}

\title{The Virtual Block Interface: A Flexible Alternative\\ to the Conventional Virtual Memory Framework} 

\author{\vspace{-12pt}\\
\protect\scalebox{0.93}{{Nastaran Hajinazar\affilETH\affilSFU}\quad%
{Pratyush Patel\affilWA}\quad%
{Minesh Patel\affilETH}\quad%
{Konstantinos Kanellopoulos\affilETH}\quad%
{Saugata Ghose\affilCMU}}\\%
\protect\scalebox{0.93}{{Rachata Ausavarungnirun\affilKING}\quad%
{Geraldo F. Oliveira\affilETH}\quad%
{Jonathan Appavoo\affilBU}\quad%
{Vivek Seshadri\affilMS}\quad%
{Onur Mutlu\affilETH\affilCMU}}\vspace{5pt}\\%
{\it\normalsize \affilETH ETH Z{\"u}rich \quad \affilSFU Simon Fraser University \quad \affilWA University of Washington \quad \affilCMU Carnegie Mellon University}\\
{\it\normalsize \affilKING King Mongkut's University of Technology North Bangkok \quad \affilBU Boston University \quad \affilMS Microsoft Research India}%
\vspace{-3pt}%
}

\maketitle
\thispagestyle{firstpage}
\pagestyle{plain}

\setstretch{0.86}
\renewcommand{\footnotelayout}{\setstretch{0.87}}

\begin{abstract}
Computers continue to diversify with respect to system designs, emerging memory technologies, and application memory demands. Unfortunately, continually adapting the conventional virtual memory framework to each possible system configuration is challenging, and often results in performance loss or requires non-trivial workarounds. 

To address these challenges, we propose a new virtual memory framework, the \sysfull (\sys). We design \sys based on the key idea that delegating memory management duties to hardware can reduce the overheads and software complexity associated with virtual memory. \sys introduces a set of variable-sized virtual blocks (VBs) to applications. Each VB is a contiguous region of the globally-visible \emph{VBI address space}, and an application can allocate each semantically meaningful unit of information (e.g., a data structure) in a separate VB. 
\sys decouples access protection from memory allocation and address translation. While the OS controls which programs have access to which VBs, dedicated hardware in the memory controller manages the physical memory allocation and address translation of the VBs. This approach enables several architectural optimizations to (1)~efficiently and flexibly cater to different and increasingly diverse system configurations, and (2)~eliminate key inefficiencies of conventional virtual memory.

We demonstrate the benefits of \sys with two important use cases: (1)~reducing the overheads of address translation (for both native execution and virtual machine environments), as \sys reduces the number of translation requests and associated memory accesses; and (2)~two heterogeneous main memory architectures, where \sys increases the effectiveness of managing fast memory regions.  For both cases, VBI significantly improves performance over conventional virtual memory.

\end{abstract}
\section{Introduction}
\label{sec:intro}

Virtual memory is a core component of modern computing systems~\cite{denning1970,fotheringham1961,kilburn1962}. Virtual memory was originally designed for systems whose memory hierarchy fit a simple two-level model of small-but-fast main memory that can be directly accessed via CPU instructions and large-but-slow external storage accessed with the help of the operating system (OS). In such a configuration, the OS can easily abstract away the underlying memory architecture details and present applications with a unified view of memory. 

However, continuing to efficiently support the conventional virtual memory framework requires significant effort due to (1)~emerging memory technologies (e.g., DRAM--NVM hybrid memories), (2)~diverse system architectures, and (3)~diverse memory requirements of modern applications.
The OS must now efficiently meet the wide range of application memory requirements that leverage the advantages offered by emerging memory architectures and new system designs while simultaneously hiding the complexity of the underlying memory and system architecture from the applications. Unfortunately, this is a difficult problem to tackle in a generalized manner. We describe three examples of challenges that arise when adapting conventional virtual memory frameworks to today's \nas{diverse system configurations}.

\textbf{Virtualized Environments.} 
\nas{In a virtual machine,} the guest OS performs virtual memory management on the emulated ``physical memory'' while the host OS performs a second round of memory management to map the emulated physical memory to the actual physical memory. This extra level of indirection results in three problems:
(1)~two-dimensional page walks~\cite{vm11, vm25, vm35, vm37, merrifield2016, pham2015},  where the number of memory accesses required to serve a TLB miss increases dramatically
(e.g., up to 24~accesses in \xeightsix with 4-level page tables);
(2)~performance loss in case of miscoordination between the guest and host OS mapping and allocation mechanisms (e.g., when the guest supports superpages, but the host does not); and
(3)~inefficiency in virtualizing increasingly complex physical memory architectures (e.g., hybrid memory systems) \nas{for the guest OS}. These problems worsen with \nas{more page table levels}~\cite{fivelevel}, and in systems that \nas{support} nested virtualization (i.e., a virtual machine running inside another)~\cite{google-nested, azure-nested}.

\textbf{Address Translation.}
In existing virtual memory frameworks, \nas{the} OS manages virtual-to-physical address mapping. However, the hardware must be able to traverse these mappings to handle memory access operations (e.g., TLB lookups). This arrangement requires using \emph{rigid} address-translation structures that are shared between and understood by both the hardware and the OS. Prior works show that many applications can benefit from flexible page tables, which cater to the application's actual memory footprint and access patterns\sg{~\cite{vm2,vm19,engler1995, kaashoek1997}}. Unfortunately, enabling such flexibility in conventional virtual memory frameworks requires more complex address translation structures \emph{every time} a new address translation approach is proposed. For example, a recent work~\cite{vm2} proposes using direct segments to accelerate big-memory applications. However, in order to support direct segments, the virtual memory contract needs to change to enable the OS to specify which regions of memory are directly mapped to physical memory. 
Despite the potential performance benefits, this approach is not easily scalable \nas{to today's increasingly diverse system architectures}.

\textbf{Memory Heterogeneity.}
\nas{Prior works propose many performance-enhancing techniques that require (1)~dynamically \emph{mapping} data to different physical memory regions according to application requirements (e.g., mapping frequently-accessed data to fast memory), and (2)~\emph{migrating} data when those requirements change (e.g., \cite{charm, diva-dram, tldram, dynsub, chang16, chang2016low, kim2018solar,clrdram,yoon2012,refree, raoux2008, li2017utility,dhiman2009pdram, ramos11,het2,zhang2009exploring,chop})}. Efficiently implementing such functionality faces two challenges. First, a customized data mapping requires the OS to be aware of microarchitectural properties of the underlying memory. Second, even if this can be achieved, the OS has low visibility into rich fine-grained runtime memory behavior information (e.g., access pattern, memory bandwidth availability), especially at the main memory level. \nas{While hardware has access to such fine-grained information, informing} the OS \emph{frequently enough} such that it can react to changes in the memory behavior of an application in a \emph{timely} manner is challenging~\cite{sim14, meswani15, ramos11, tumanov13,banshee}.

A wide body of research (e.g., \cite{vm1,vm2,vm3,karakostas2015,vm5,vm6,pichai2014,vm8,mask,vm9,vm10,vm11,vm12,vm13,vm14,pham2014,vm16,pham2015,vm18,vm19,vm20,vm21,vm22,vm23,vm24,vm25,vm26,vm27,vm28,vm29,vm30,vm31,vm32,vm33,vm34,vm35,vm36,vm37,vm38,vm39,vm40,vm41,vm42,meswani15,het9,het10,sim14,het12,mitosis-asplos20,elastic-cuckoo-asplos20,meza2013}) proposes mechanisms to alleviate the overheads of conventional memory allocation and address translation by exploiting specific trends observed in modern systems (e.g., the behavior of emerging applications). Despite notable improvements, these solutions have two major shortcomings. First, these solutions mainly exploit specific system or workload characteristics and, thus, are applicable to a limited set of problems or applications. Second, each solution requires specialized and not necessarily compatible changes to both the OS and hardware. Therefore, implementing all of these proposals at the same time in a system is a daunting prospect.

\textbf{Our goal} in this work is to design \emph{a general-purpose alternative virtual memory framework that naturally supports and better extracts performance from a wide variety of new system configurations, while still providing the key features of conventional virtual memory frameworks}. To this end, we propose the \sysfull (\sys), an alternative approach to memory virtualization that is inspired by the logical block abstraction used by solid-state drives to hide the underlying device details from the rest of the system. In a similar way, we envision the memory controller as the primary provider of an abstract interface that hides the details of the underlying physical memory architecture, including the physical addresses of the memory locations. 

VBI is based on three guiding principles. First, \emph{programs should be allowed to choose the \emph{size} of their virtual address space}, to mitigate translation overheads associated with very large virtual address spaces. Second, \emph{address translation should be decoupled from memory protection}, since they are logically separate and need not be managed at the same granularity by the same structures. Third, \emph{\sgii{software should be allowed} to communicate semantic information about application data to the hardware}, so that the hardware can \sgii{more intelligently manage} the underlying hardware resources.

\sys introduces a \emph{globally-visible} address space called the \emph{\sys Address Space}, that consists of a large set of \emph{virtual blocks (VBs)} of different sizes. \nasi{For any semantically meaningful
unit of information (e.g., a data structure, a shared library)}, the program can choose a VB of appropriate size, and tag the VB with properties that describe the contents of the VB. \textbf{The key idea} of \sys is to delegate physical memory allocation and address translation to a hardware-based \mtlfull (MTL) at the memory controller. This idea is enabled by the fact that the globally-visible \sys address space provides \sys with system-wide unique \emph{\sys addresses} that can be \emph{directly} used by on-chip caches without requiring address translation. In \sys, the OS no longer needs to manage address translation and memory allocation for the physical memory devices. Instead, the OS (1)~retains full control over access protection by controlling which programs have access to which virtual blocks, and (2)~uses VB properties to communicate the data's memory requirements (e.g., latency sensitivity) and characteristics (e.g., access pattern) to the memory controller.

\begin{figure}[h]
    \centering
    \begin{subfigure}[b]{0.455\columnwidth}
      \centering
      \includegraphics[width=\textwidth, trim=0 35 266 0, clip]{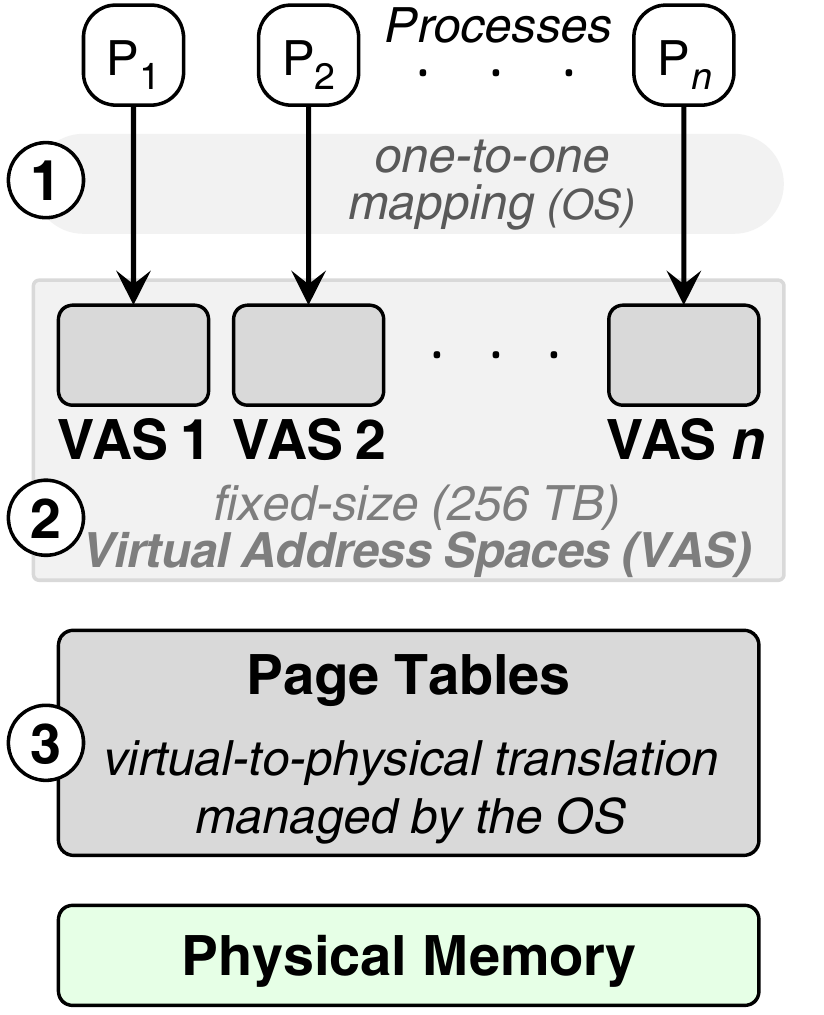}
      \caption{Intel x86-64}
    \end{subfigure}%
    \hfill\vline\hfill%
    \begin{subfigure}[b]{0.455\columnwidth}
      \centering
      \includegraphics[width=\textwidth, trim=266 35 0 0, clip]{figures/vbi-intro.pdf}
      \caption{Virtual Block Interface}
    \end{subfigure}%
    \caption{Virtual memory management in \xeightsix and in \sys.}
    \label{fig:vbi-intro}
\end{figure}

Figure~\ref{fig:vbi-intro} illustrates the differences between virtual memory management in state-of-the-art production Intel \xeightsix systems~\cite{intelx86manual} and in \sys. In \xeightsix (Figure~\ref{fig:vbi-intro}a), the OS manages a single private virtual address space (VAS) for each process (\inhollowcircle{1}), providing each process with a fixed-size 256 TB VAS irrespective of the actual memory requirements of the process (\inhollowcircle{2}). The OS uses a set of page tables, one per process, to define how each VAS maps to physical memory (\inhollowcircle{3}). In contrast, \sys (Figure~\ref{fig:vbi-intro}b) makes \emph{all} virtual blocks (VBs) visible to \emph{all} processes, and the OS controls which processes can access which VBs (\incircle{1}). Therefore, a process' total virtual address space is defined by which VBs are attached to it, i.e., by the process' actual memory needs (\incircle{2}). In \sys, the \mtl has full control over mapping of data from each VB to physical memory, invisibly to the system software (\incircle{3}).

\sys seamlessly and efficiently supports important optimizations that improve overall system performance, including: (1)~enabling benefits akin to using virtually-indexed virtually-tagged (VIVT) caches (e.g., reduced address translation overhead), (2)~eliminating two-dimensional page table walks in virtual machine environments, (3)~delaying physical memory allocation until the first dirty last-level cache line eviction, and (4)~flexibly supporting different virtual-to-physical address translation structures for different memory regions. \pointer{\secref{sec:optimizations}} describes these optimizations in detail.

We evaluate VBI for two important and emerging use-cases. First, we demonstrate that \sys significantly reduces the address translation overhead both for \emph{natively-running programs} and for programs running inside a virtual machine (\emph{VM programs}). Quantitative evaluations using workloads from SPEC CPU 2006~\cite{spec2006}, SPEC CPU 2017~\cite{spec2017}, TailBench~\cite{tailbench}, and Graph 500~\cite{graph500} show that a simplified version of \sys that maps VBs using 4~KB granularity only improves the performance of native programs by 2.18$\times$ and VM programs by 3.8$\times$. Even when enabling support for large pages for \emph{all data}, which significantly lowers translation overheads, \sys improves performance by 77\% for native programs and 89\% for VM programs. Second, we demonstrate that \sys significantly improves the performance of heterogeneous memory architectures by evaluating two heterogeneous memory systems (PCM--DRAM~\cite{ramos11} and Tiered-Latency-DRAM~\cite{tldram}). We show that \sys, by intelligently mapping frequently-accessed data to the low-latency region of memory, improves overall performance of these two systems by 33\% and 21\% respectively, compared to systems that employ a heterogeneity-unaware data mapping scheme. \pointer{\secref{sec:methodology2}} describes our methodology, results, and insights from these evaluations.

We make the following key contributions:
\begin{itemize}
    \item To our knowledge, this is the first work to propose a virtual memory framework that relieves the OS of explicit physical memory management and delegates this duty to the hardware, i.e., the memory controller. 
    \item We propose VBI, a new virtual memory framework that efficiently enables memory-controller-based memory management by exposing a purely virtual memory interface to applications, the OS, and the hardware caches. \sys naturally and seamlessly supports several optimizations (e.g., low-cost page walks in virtual machines, purely virtual caches, delayed physical memory allocation), and integrates well with a wide range of system designs.
    \item We provide a detailed reference implementation of \sys, including required changes to the user applications, system software, ISA, and hardware.
    \item We quantitatively evaluate \sys using two concrete use cases: (1)~address translation improvements for native execution and virtual machines, and (2)~two different heterogeneous memory architectures. Our evaluations show that \sys significantly improves performance in both use cases.
\end{itemize}

\section{\nas{Design Principles}}
\label{sec:design-principles}

To minimize performance and complexity overheads of memory virtualization, our virtual memory framework is grounded on three key design principles.

\paragraph{Appropriately-Sized Virtual Address Spaces.}
The virtual memory framework should \emph{allow each application to have control over the size of its virtual address space}. The majority of applications far underutilize the large virtual address space offered by modern architectures (e.g., 256~TB in Intel \xeightsix). Even demanding applications such as databases~\cite{db1,db2,db3,db4,db5,db6} and caching servers~\cite{memcache,memcached1} are cognizant of the amount of available physical memory and of the size of virtual memory they need. Unfortunately, a larger virtual address space results in larger or deeper page tables (i.e., page tables with more levels). A larger page table increases TLB contention, while a deeper page table requires a greater number of page table accesses to retrieve the physical address for each TLB miss.  In both cases, the address translation overhead increases. Therefore, allowing applications to choose an appropriately-sized virtual address space based on their actual needs, avoids the higher translation overheads associated with a larger address space.

\paragraph{Decoupling Address Translation from Access Protection.}
The virtual memory framework should \emph{decouple address translation from access protection checks}, as the two have inherently different characteristics. While address translation is typically performed at page granularity, protection information is typically the same for an entire data structure, which can span multiple pages. Moreover, protection information is purely a function of the virtual address, and does not require address translation. However, existing systems store both translation and protection information for each virtual page as part of the page table. Decoupling address translation from protection checking can enable opportunities to remove address translation from the critical path of an access protection check, deferring the translation until physical memory \mhp{\emph{must}} be accessed, thereby lowering the performance overheads of virtual memory.

\paragraph{Better Partitioning of Duties Between Software and Hardware.}
The virtual memory framework should \emph{allow software to easily communicate semantic information about application data to hardware and allow hardware to manage the physical} memory resources. Different pieces of program data have different performance characteristics (latency, bandwidth, and parallelism), and have other inherent properties (e.g., compressibility, persistence) at the software level. As highlighted by recent work~\cite{xmem,vijaykumar2018}, while software is aware of this semantic information, the hardware is privy to \mhp{fine-grained} dynamic runtime information \mhp{(e.g., memory access} behavior, phase changes, \mhp{memory} bandwidth availability) that can enable vastly more intelligent management of the underlying hardware resources (e.g., better data mapping, migration, and scheduling decisions). Therefore, conveying semantic information to the hardware (i.e., memory controller) that manages the \mhp{physical} memory resources can enable a host of new optimization opportunities. 
\section{Virtual Block Interface\mhp{:} Overview}
\label{sec:overview}

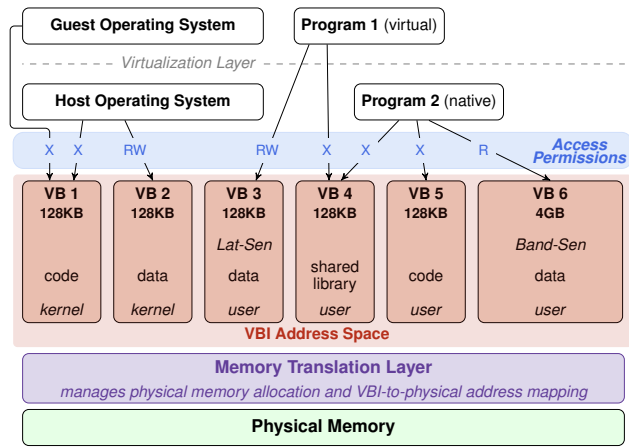
\begin{figure}[h]
    \centering
    \scalebox{0.8}{\begin{tikzpicture}[>=stealth',font=\small\sffamily]

  \tikzset{block/.style={draw,rounded corners=3pt,minimum height=0.6cm,align=center}};
  \tikzset{acllabel/.style={fill=CornflowerBlue!20,text=RoyalBlue,font=\footnotesize\sffamily}};
  \tikzset{acl/.style={->,rounded corners=2pt}};
  \tikzset{mmap/.style={->,rounded corners=2pt}};
  \tikzset{vb/.style={draw,rounded corners=3pt,minimum height=1.2cm,minimum width=1.3cm,align=center,fill=BrickRed!15}};
  \tikzset{client/.style={draw,rounded corners=3pt,minimum height=1cm,align=center,fill=black!20}};
  
  \node (mml) [block, minimum width=10cm,minimum height=0.8cm,RoyalPurple, fill=RoyalPurple!15] {\normalsize\textbf{Memory Translation Layer}\\\emph{manages physical memory allocation and VBI-to-physical address mapping\vspace{-2pt}}};
  
  \node (pmem1) [block, minimum width=10cm,minimum height=0.6cm, fill=green!10, yshift=-4mm] at (mml.south) {\normalsize\textbf{Physical Memory}};

  \node (vb1) at (mml.north west) [anchor=south west,yshift=5mm,vb] {\textbf{VB 1}\\[-0.5ex]\textbf{\footnotesize{128KB}}\\[1ex]\\[1ex]code\\[-1.5ex]\\\emph{kernel}};
  \node (vb2) at (vb1.east) [anchor=west,xshift=2mm,vb] {\textbf{VB 2}\\[-0.5ex]\textbf{\footnotesize{128KB}}\\[1ex]\\[1ex]data\\[-1.5ex]\\\emph{kernel}};
  \node (vb4) at (vb2.east) [anchor=west,xshift=2mm,minimum width=1.75cm,vb] {\textbf{VB 3}\\[-0.5ex]\textbf{\footnotesize{128KB}}\\[1ex]\emph{Lat-Sen}\\[1ex]data\\[-1.5ex]\\\emph{user}};
  \node (vb5) at (vb4.east) [anchor=west,xshift=2mm,vb] {\textbf{VB 4}\\[-0.5ex]\textbf{\footnotesize{128KB}}\\[0.9ex]\\shared\\[-0.8ex]library\\[0.4ex]\emph{user}};
  \node (vb6) at (vb5.east) [anchor=west,xshift=2mm,vb] {\textbf{VB 5}\\[-0.5ex]\textbf{\footnotesize{128KB}}\\[1ex]\\[1ex]code\\[-1.5ex]\\\emph{user}};
  \node (vb7) at (vb6.east) [anchor=west,xshift=2mm,vb,minimum width=2.4cm] {\textbf{VB 6}\\[-0.5ex]\textbf{\footnotesize{4GB}}\\[1ex]\emph{Band-Sen}\\[1ex]data\\[-1.5ex]\\\emph{user}};
  
       \node at (vb5.south) [yshift=-2mm, xshift=-3mm,text=BrickRed]{\textbf{VBI Address Space}};
  
  \node (hostos) at (vb1.north west) [anchor=south west,minimum width=4cm,yshift=10mm,block] {\textbf{Host Operating System}};
  \node (p1) at (hostos.east) [anchor=west,xshift=15mm,block] {\textbf{Program 2} (native)};

  \node (vl) at (hostos.north west) [anchor=south west,yshift=2mm,minimum width=10cm,align=center] {};
  \node (vllabel) at (vl.west) [anchor=west,xshift=15mm,black!50] {\emph{Virtualization Layer}};
  \draw [dashed,black!50] (vllabel.west) -- (vl.west);
  \draw [dashed,black!50] (vllabel.east) -- (vl.east);

  \node (guestos) at (vl.north west) [anchor=south west,yshift=2mm,minimum width=4cm,block] {\textbf{Guest Operating System}};
  \node (p2) at (guestos.east) [anchor=west,xshift=5mm,block] {\textbf{Program 1} (virtual)};
  
  \draw [rounded corners=5pt,color=CornflowerBlue!30,fill=CornflowerBlue!20] ([yshift=2mm,xshift=-1.5mm]vb1.north west) rectangle ([yshift=8mm,xshift=1.5mm]vb7.north east);

     \fill [BrickRed,fill opacity=0.15,rounded corners=2pt] ([yshift=1mm,xshift=-1.5mm]vb1.north west) rectangle ([yshift=-4mm,xshift=1.5mm]vb7.south east);

  \draw [acl] ([xshift=-10mm]hostos.south) -- ([xshift=2mm]vb1.north);
  \draw [acl] ([xshift=-3mm]hostos.south) -- ([xshift=0mm]vb2.north);
  \draw [acl] (guestos.west) -- ++(-2mm,0) -- ++(0,-1.7cm) -| ([xshift=-2mm]vb1.north);

  \draw [acl] (p1.south) -- (vb7.north);
  \draw [acl] ([xshift=-2mm]p1.south) -- (vb6.north);
  \draw [acl] ([xshift=-5mm]p1.south) -- ([xshift=1mm]vb5.north);

  \draw [acl] ([xshift=-10mm]p2.south) -- ([xshift=2mm]vb4.north);
  \draw [acl] ([xshift=-7mm]p2.south) -- ([xshift=-1mm]vb5.north);

  \node (l1) [acllabel] at ([yshift=5mm,xshift=-2mm]vb1.north) {X};
  \node (l2) [acllabel,anchor=west] at ([xshift=0.7mm]l1.east) {X};
  \node (l2) [acllabel,anchor=west] at ([xshift=9mm]l1.east) {RW};
  \node (l3) [acllabel,anchor=west] at ([xshift=31mm]l1.east) {RW};
  \node (l4) [acllabel,anchor=west] at ([xshift=42mm]l1.east) {X};
  \node (l4) [acllabel,anchor=west] at ([xshift=48.5mm]l1.east) {X};
  \node (l4) [acllabel,anchor=west] at ([xshift=57.5mm]l1.east) {X};
  \node (l4) [acllabel,anchor=west] at ([xshift=68mm]l1.east) {R};
  
    \node (label) at (l1.east) [yshift=1mm,xshift=86mm,text=RoyalBlue,align=center]{\textbf{\emph{Access}}};
    \node at (label.south) [text=RoyalBlue,align=center]{\textbf{\emph{Permissions}}};

\end{tikzpicture}}
    \caption{Overview of \sys. \emph{Lat-Sen} and \emph{Band-Sen} represent latency-sensitive and bandwidth-sensitive,
respectively.}
    \label{fig:design-overview}
    \vspace{-5pt}
\end{figure}

\nas{Figure~\ref{fig:design-overview} shows an overview of \sys.} There are three major aspects of the \sys design: (1)~the \mhp{\sys address space}, (2)~\sys access permissions, and (3)~the \mtlfull. We \mhp{first} describe these aspects in detail (\pointer{\secref{sec:add-space}}--\pointer{\secref{sec:mtl-overview}}). Next, \nasi{we explain the implementation of key OS functionalities in \sys}~\mhp{(\pointer{\secref{sec:os-functions}}). Finally, we discuss some of the \mhp{key} optimizations that \sys enables \mhp{(\pointer{\secref{sec:optimizations}})}}.

\subsection{\sys Address Space}
\label{sec:add-space}

Unlike most existing architectures wherein each process has its own virtual address space, virtual memory in \sys is a single, globally-visible address space called the \emph{\sys Address Space}. As shown in Figure~\ref{fig:design-overview}, the \sys Address Space consists of a finite set of \emph{Virtual Blocks} (VBs). Each VB is a contiguous region of VBI address space that does not overlap with any other VB. \nasi{Each VB contains a semantically meaningful
unit of information (e.g., a data structure, a shared library)} and is associated with (1)~a system-wide unique ID, (2)~a specific size (chosen from a set of pre-determined size classes), and (3)~a set of properties that specify the semantics of the content of the VB and its desired characteristics. For example, in the figure, \texttt{VB~1} indicates the VB with ID 1; its size is 128~KB, and it contains code that is accessible only to the kernel. On the other hand, \texttt{VB~6} is the VB with ID 6; its size is 4~GB, and it contains data that is bandwidth-sensitive. 
In contrast to conventional systems, where the mapping from the process' \nasi{virtual-to-physical address space} is stored in a per-process page table~\cite{intelx86manual}, \sys maintains the VBI-to-physical address mapping information of each VB in a \emph{separate} translation structure. This approach enables \sys to flexibly tune the type of translation structure for each VB to the characteristics of the VB~(as described in \secref{sec:flexible-translations}). \sys stores the above information and a pointer to the translation structure of each VB in a set of \emph{VB Info Tables} (VITs; described in \pointer{\secref{sec:vit}}).

\subsection{\sys Access {Permissions}}
\label{sec:access-semantics}

As the \sys Address Space is global, all VBs in the system are \emph{visible} to all processes. However, a program can \emph{access} data within a VB \emph{only if} it is attached to the VB with appropriate permissions. In Figure~\ref{fig:design-overview}, \texttt{Program 2} can only execute from \texttt{VB 4} or \texttt{VB 5}, only read from \texttt{VB 6}, and cannot access \texttt{VB 3} at all; \texttt{Program 1} and \texttt{Program 2} both share \texttt{VB 4}. For each process, \sys maintains information about the set of VBs attached to the process in an {OS-managed} {per-process} table called \mhp{the} \emph{Client--VB Table} (CVT) (described in \pointer{\secref{sec:client}}). \sys provides the OS with a set of instructions with which the OS can control which processes have what type of access permissions to which VBs. On each memory access, the processor checks the CVT to ensure that the program has the necessary permission to perform the access. With this approach, \sys \emph{decouples protection checks from address translation}, which allows it to defer the address translation to the memory controller where the physical address is required to access \mhp{main} memory.

\subsection{\mtlfull}
\label{sec:mtl-overview}

In \sys, to access a piece of data, a program must specify the ID of the VB that contains the data and the offset of the data within the VB. Since the {ID of the VB} is unique system-wide, {the} combination of {the ID} and offset points to the address of a specific byte of data in the \sys address space. We call this address the \emph{\sys address}. As the \sys address space is globally visible, similar to the physical address in existing architectures, the \sys address points to a unique piece of data in the system. As a result, \sys uses the VBI address \emph{directly} (i.e., without requiring address translation) to locate data within the on-chip caches without worrying about the complexity of homonyms and synonyms~\cite{cekleov1997a, cekleov1997b, jacob1998}{, which cannot exist in VBI (see \pointer{\secref{sec:optimizations}})}. Address translation is required only when an access misses in all levels of on-chip caches.

{To perform address translation,} \sys uses the \mtlfull (\mtl). The \mtl, implemented in the memory controller with an interface to the system software, manages both allocation of physical memory to VBs and VBI-to-physical address translation (relieving the OS of these duties). {Memory-controller-based memory management} enables a number of performance optimizations (e.g., avoiding 2D page walks in virtual machines, flexible address translation structures), which we describe in \pointer{\secref{sec:optimizations}}.

\subsection{Implementing Key OS Functionalities}
\label{sec:os-functions}

\sys allows the system to efficiently implement existing OS functionalities. In this section, we describe five key functionalities {and how VBI enables them}.

\paragraph{Physical Memory Capacity Management.} In \sys, the \mtl allocates physical memory for VBs as and when required. To handle situations when the \mtl runs out of physical memory, \sys provides two system calls that allow the \mtl to move data from physical memory to the backing store and vice versa. The \mtl maintains information about swapped-out data as part of the VB's translation structures.

\paragraph{Data Protection.} The goal of data protection is to prevent a malicious program from accessing kernel data or private data of other programs. In \sys, the OS ensures such protection by appropriately setting the permissions with which each process can access different VBs. Before each \sgii{memory access}, the CPU checks if the executing thread has appropriate access permissions to the corresponding VB (\pointer{\secref{sec:load}}).

\paragraph{Inter-Process Data Sharing (True Sharing).} When two processes share data (e.g., via pipes), both processes have a coherent view of the shared memory, i.e., modifications made by one process should be visible to the other process. In \sys, the OS supports such \emph{true} sharing by granting both processes permission to access the VB containing the shared data.

\paragraph{Data Deduplication (Copy-on-Write Sharing).} In most modern systems, the OS reduces redundancy in physical memory by mapping virtual pages {containing} the \emph{same} data to the same physical page. On a write to one of the virtual pages, the OS copies the data {to a new physical page,} and remaps the written virtual page to the new physical page before performing the write. In \sys, the \mtl performs data deduplication when a VB is cloned by sharing both translation structures and data pages between the two VBs (\pointer{\secref{sec:pop-interaction}}), and using the copy-on-write mechanism to ensure consistency.  

\paragraph{Memory-Mapped Files.} To support memory-mapped files, existing systems map a region of the virtual address space to a file in storage, and loads/stores to that region are used  to access/update the file content. \sys naturally supports memory-mapped files as the OS simply associates the file to a VB of appropriate size. An offset within the VB maps to the same offset within the file. The \mtl uses the same system calls used to manage physical memory capacity (described under \emph{Physical Memory Capacity Management} above) to move data between the VB in memory and the file in storage.

\subsection{ Optimizations Supported by \sys}
\label{sec:optimizations}

In this section, we describe four {key} optimizations that the \sys design enables.

\paragraph{Virtually-Indexed Virtually-Tagged Caches.}
Using {fully-virtual (i.e., VIVT) caches} enables the system to delay address translation and {reduce accesses to translation structures} such as the {TLBs}. However, most modern architectures do not support VIVT caches due to two main reasons. First, handling homonyms (i.e., where the same virtual address maps to multiple physical addresses) and synonyms (i.e., where multiple virtual addresses map to the same physical address) introduces complexity to the system~\cite{cekleov1997a, cekleov1997b, jacob1998}. Second, although address translation is not required to access VIVT caches, the access permission check required prior to the cache access still necessitates accessing the TLB and can induce a page table walk on a TLB miss. This is due to the fact that the protection bits are stored as part of the page table entry for each page in current systems. VBI avoids both of these problems. 

First, VBI addresses are unique system-wide, eliminating the possibility of homonyms. Furthermore, since VBs do not overlap, each VBI address appears in \emph{at most one} VB, {avoiding the possibility of} synonyms. In case of true sharing (\pointer{\secref{sec:os-functions}}), different processes are attached to the same VB. Therefore, the VBI address that each process uses to access the shared region refers to the \emph{same} VB. In case of copy-on-write sharing, where the MTL may map two VBI addresses to the same physical memory for deduplication, the MTL creates a new copy of the data before any write to either address. {Thus, neither form of sharing can lead to synonyms.} As a result, by using VBI addresses directly to access on-chip caches, \sys achieves benefits akin to VIVT caches without the complexity of dealing with synonyms and homonyms. Additionally, since the VBI address acts as a system-wide single point of reference for the data that it refers to, all coherence-related requests can use VBI addresses without introducing any ambiguity.

Second, \sys decouples protection checks from address translation, by storing protection and address translation information in \emph{separate} sets of tables and delegating access permission management to the OS{, avoiding the need to access translation structures for protection purposes (as done in existing systems).}

\paragraph{Avoiding 2D Page Walks in Virtual Machines.}
In \sys, once a process inside a VM attaches itself to a VB (with the help of the host and guest OSes), any memory access from the VM directly {uses} a VBI address. As described in \pointer{\secref{sec:mtl-overview}}, this address is directly used to address the on-chip caches. In case of an LLC miss, the \mtl translates the VBI address to physical address. As a result, unlike existing systems, {address translation for a VM under VBI is no different from that for a host}, enabling significant performance improvements. We expect these benefits to further increase in systems supporting nested virtualization~\cite{google-nested,azure-nested}. \pointer{\secref{sec:virtual-machines}} discusses the implementation of \sys in virtualized environments.

\paragraph{Delayed Physical Memory Allocation.}
As \sys uses VBI addresses to access all on-chip caches, it is no longer necessary for a cache line to be backed by physical memory \emph{before} it can be accessed. This enables the opportunity to delay physical memory allocation for a VB (or a region of a VB) until a dirty cache line from the VB is evicted from the last-level cache. Delayed allocation has three benefits. First, the allocation process is removed from the critical path of execution, as cache line evictions are not on the critical path. Second, for VBs that never leave the cache during the lifetime of the VB (likely more common with growing cache sizes in modern hardware), \sys avoids physical memory allocation altogether. Third, \nas{when using delayed physical memory allocation, }for an access to a region with no physical memory allocated yet, \sys simply returns a zero cache line, thereby avoiding \emph{both} address translation and a main memory access, which improves performance. \pointer{\secref{sec:delayed-allocation}} describes the implementation of delayed physical memory allocation in \sys.

\paragraph{Flexible Address Translation Structures.}
A recent work~\cite{vm19} shows that different data structures benefit from different types of address translation structures depending on their data layout and access patterns. However, since in {conventional virtual memory, the} hardware needs to read the OS-managed page tables to perform page table walks, the structure of the page table needs to be understood by both the hardware and OS, thereby limiting the flexibility of the page table structure. In contrast, in \sys, the \mtl is the \emph{only} component that manages and accesses translation structures. Therefore, the constraint of sharing address translation structures with the  OS is relaxed, providing \sys with more flexibility in employing different types of translation structures \mhp{in the MTL}. Accordingly, \sys maintains a separate translation structure for each VB, and can tune {it} to suit the properties of the VB (e.g., multi-level tables for large VBs or those with many sparsely-allocated regions, and single-level tables for small VBs or those with many \nas{large} contiguously-allocated regions). This optimization reduces the number of memory accesses necessary to serve a TLB miss.

\section{\sys: Detailed Design}
\label{sec:detailed}

In this section, we present the detailed design and a reference implementation of the \sysfull. {We describe (1)}~the components architecturally exposed by \sys to the rest of the system (\pointer{\secref{sec:arch-interface}}){, (2)}~the life-cycle of {allocated memory} (\pointer{\secref{sec:ds-life-cycle}}){, (3)}~the interactions between the processor, OS, and the process in \sys (\pointer{\secref{sec:pop-interaction}}){, and (4)}~the operation of the \mtlfull in detail (\pointer{\secref{sec:mtl}}). 

\subsection{Architectural Components}
\label{sec:arch-interface}

\sys exposes two architectural components to the rest of the system that form the contract between hardware and software: (1)~{virtual blocks}, and (2)~\emph{memory clients}.

\subsubsection{{Virtual  Blocks (VBs)}}
\label{sec:vb-detail}

\begin{figure}[h]
  \centering
  \begin{tikzpicture}[inner sep=1pt,outer sep=0pt,>=stealth']

\draw [rounded corners=2pt] (0,0) rectangle (8,0.7);
\draw (4.4,0) -- ++(0,0.7);
\draw[dotted] (1.4,0) -- ++(0,0.7);

\node at (2.2, 0.5) {\small\texttt{VBUID}};
\node at (0.7, 0.2) {\footnotesize\color{Gray}\texttt{\emph{SizeID}}};
\node at (2.8, 0.2) {\footnotesize\color{Gray}\texttt{\emph{VBID}}};
\node at (6.2, 0.35) {\small\texttt{offset}};

\draw[->] (1.75, 0.5) -- ++ (-1.7,0);
\draw[->] (2.65, 0.5) -- ++ (1.7,0);

\end{tikzpicture}
  \caption{Components of a VBI address.}
  \label{fig:vbi-address}
\end{figure}
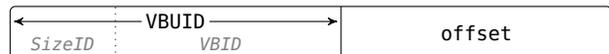

The \mhp{\sys address space} in \sys is characterized by three parameters: (1)~the size of the address space, which is determined by the bit width of the processor's address bus (64 in our implementation); (2)~the number of VB size classes (8 in our implementation); and (3)~the list of size classes (4~KB, 128~KB, 4~MB, 128~MB, 4~GB, 128~GB, 4~TB, and 128~TB). \nas{Each size class in \sys is associated with an ID (\texttt{SizeID}), and each VB is assigned an ID \emph{within its size class} (\texttt{VBID}).
Every VB is identified system-wide by its \emph{VBI unique ID} (\texttt{VBUID}), which is the concatenation of \texttt{SizeID} and \texttt{VBID}. As shown in Figure~\ref{fig:vbi-address}, \sys constructs a \mhp{\emph{\sys address}} using two components:
(1)~\texttt{VBUID}, and (2)~the offset of the addressed data within the VB. In our implementation, \texttt{SizeID} uses three bits to represent each of our eight possible size classes.
The remaining address bits are split between 
\texttt{VBID} and the offset. The precise number of bits required for the offset is determined by the size of the VB, and the remaining bits are used for \texttt{VBID}.} For example, the 4~KB size class in our implementation uses 12~bits for the offset, leaving 49~bits for \texttt{VBID}, i.e., 2$^{49}$ VBs of size 4~KB. In contrast, the 128~TB size class uses 47 bits for the offset, leaving 14~bits for \texttt{VBID}, i.e., 2$^{14}$ VBs of size 128~TB.

As \pointer{\secref{sec:overview}} describes, \sys associates each VB with a set of flags that characterize the contents of the VB (e.g, \texttt{code}, \texttt{read-only}, \texttt{kernel}, \texttt{compressible}, \texttt{persistent}). In addition to these flags,  software may also provide hints to describe the memory behavior of the data that the VB contains (e.g., latency sensitivity, bandwidth sensitivity, {compressibility, error tolerance}). Prior work extensively studies a set of useful properties{~\cite{xmem,vijaykumar2018,yixin2014,yu2017}}. \mhp{Software} specifies these properties via a bitvector \mhp{that is defined as part of the ISA specification}. \sys maintains the flags and the \mhp{software-provided} hints as a \emph{property bitvector}.

For each VB {in the system}, \sys stores (1)~\mhp{an \emph{enable} bit to describe whether the VB is currently assigned to any process}, (2)~the property bitvector, (3)~the number of processes attached to the VB (i.e., a reference count), \nas{(4)~the type of VBI-to-physical address translation structure being used for the VB, and} (5)~a pointer to the VB's address translation structure. All {of this information is} stored as {an entry in} the VB Info Tables (\pointer{\secref{sec:vit}}).

\subsubsection{Memory Clients}
\label{sec:client}

Similar to \sg{address space identifiers~\cite{ahearn1973}} in existing architectures, \sys introduces the notion of \emph{memory client} to communicate the concept of a process in \sys. \nasi{A memory client refers to any entity that needs to allocate and use memory, such as the OS itself, and any process running on the system (natively or inside a virtual machine).} In order to track the
permissions with which a client can access different VBs,
each client in VBI is assigned a unique ID to identify
the client system-wide. During execution, \sys tags each core with the client ID of the process currently running on it.

As \pointer{\secref{sec:overview}} discusses, the set of VBs that a client can access and their associated permissions are stored in a per-client table called the \emph{Client--VB Table} (CVT). Each entry in the CVT contains 
(1)~a valid bit, 
(2)~\texttt{VBUID} of the VB, and
(3)~a three-bit field representing the read-write-execute permissions (RWX) with which the client can access that VB. For each memory access, the processor checks the CVT to ensure that the client has appropriate access to the VB. The OS implicitly manages the CVTs using the following two new instructions:

\begin{figure}[h!]\small
  \centering
  \begin{tikzpicture}
    \node (attach) [draw,rounded corners=2pt,inner sep=5pt] {\tt \textcolor{blue}{\attachvb} CID, VBUID, RWX};
    \node (detach) at (attach.east) [draw,rounded corners=2pt,inner sep=5pt,anchor=west,xshift=8mm] {\tt \textcolor{blue}{\detachvb} CID, VBUID};
  \end{tikzpicture}
\end{figure}

The \attachvb instruction adds an entry for VB \texttt{VBUID} in the CVT of client \texttt{CID} with the specified \texttt{RWX} permissions (either by replacing an invalid entry in the CVT, or being inserted at the end of the CVT). This instruction returns the index of the CVT entry to the OS {and increments the reference count of the VB (stored in the VIT entry of the VB; see \pointer{\secref{sec:vit}})}. The \detachvb instruction resets the valid bit of the entry corresponding to VB \texttt{VBUID} in the CVT of client \texttt{CID} and decrements the reference count of the VB.

The processor maintains the location and size of the CVT for each client in a reserved region of physical memory. As clients are visible to both the hardware and the software, the number of clients is an architectural parameter determined at design time and exposed to the OS. In our implementation, we use 16-bit client IDs (supporting 2$^{16}$ clients).

\subsection{Life Cycle of \nas{Allocated Memory}}
\label{sec:ds-life-cycle}

\begin{figure*}[t]
  \centering
  \includegraphics[width=\textwidth]{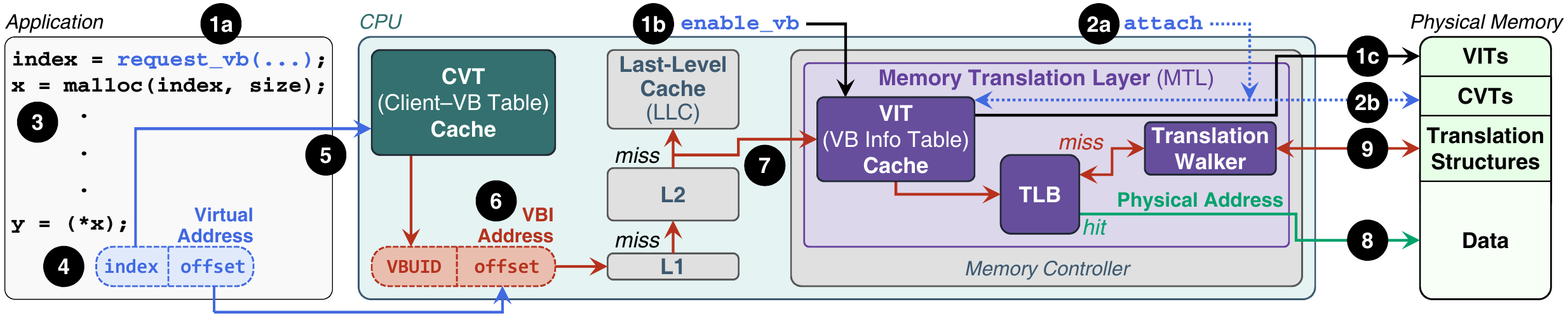}
  \vspace{-15pt}
  \caption{Reference microarchitectural implementation of the Virtual Block Interface.}
  \label{fig:vbi-uarch}
\end{figure*}

In this section, we describe the phases in the life cycle of \nas{dynamically-allocated memory}: memory allocation, address specification, data access, and deallocation. Figure~\ref{fig:vbi-uarch} shows this flow in detail, including the hardware components that aid \sys in efficiently executing memory operations. In \pointer{\secref{sec:pop-interaction}}, we discuss how VBI \nas{manages code}, shared libraries, {static data}, and the life cycle of an entire process.

\nas{When} a program needs to allocate memory for a new data structure, it first requests a new VB from the OS. For this purpose, we introduce a new system call, \requestvas. The program invokes \requestvas with two parameters: (1)~the \emph{expected} size of the data structure, and (2)~a bitvector of the desired properties for the data structure \sg{(\incircle{1a} in Figure~\ref{fig:vbi-uarch})}.

In response, the OS first scans the VB Info Table to identify the smallest free VB that can accommodate the data structure. 
\sg{The OS then uses the \enablevb instruction (\incircle{1b}) to inform the MTL that the VB is now \nas{enabled}}. The \enablevb instruction takes the \texttt{VBUID} of the VB to be enabled along with the properties bitvector as arguments. Upon executing this instruction, the MTL updates the entry for the VB in the VB Info Table to reflect that it is now enabled with the appropriate properties (\incircle{1c}).

\begin{figure}[h!]\small
  \centering
  \begin{tikzpicture}
    \node (enable) [draw,rounded corners=2pt,inner sep=5pt] {\tt \textcolor{blue}{\enablevb} VBUID, props};
  \end{tikzpicture}
  \vspace{-3pt}
\end{figure}

\subsubsection{Dynamic Memory Allocation}
\label{sec:dynamic_alloc}

After enabling the VB, the OS \sg{uses the \attachvb instruction (\incircle{2a}) to add the VB to the CVT of the calling process {and increment the \sgii{VB's} reference count in its VIT entry} (\incircle{2b}; \pointer{\secref{sec:client}}).} The OS then returns the index of the \mhp{newly-added} CVT entry as the return value of the \requestvas system call (stored as \texttt{index} in the application code \mhp{example} of Figure~\ref{fig:vbi-uarch}). This \texttt{index} serves as a pointer to the VB. As we discuss in \mhp{\pointer{\secref{subsubsec:addr_spec}}}, the program \sgii{uses} this index to specify virtual addresses to the processor.

After the VB is attached to the process, the process can access any location within the VB with the appropriate permissions. It can also dynamically manage memory inside the VB using modified versions of \texttt{malloc} and \texttt{free} that take the CVT entry \texttt{index} as an additional argument \sg{(\incircle{3})}. During execution, it is possible that the process runs out of memory within a VB (\nas{e.g., }due to an incorrect estimate of the expected size of the data structure). In such a case, \sys allows automatic promotion of the \nas{allocated data} to a VB of a larger size class. \pointer{\secref{sec:vb-promotion}} discusses VB promotion in detail.

\subsubsection{Address Specification}
\label{subsubsec:addr_spec}

In order to access data inside a VB, the process generates a two-part virtual address in the format of $\{${\tt \cvt index, offset}$\}$. The \texttt{CVT index} specifies the CVT entry that points to the corresponding VB, and the \texttt{offset} is the location of the data inside the VB. Accessing the data indirectly through the \texttt{CVT index} as opposed to direcly using the VBI address allows \sys to \nas{not require relocatable code and maintain the validity of the pointers (i.e., virtual addresses) within a VB when migrating/copying the content of a VB to another VB. \mhp{With CVT indirection, VBI can} seamlessly \mhp{migrate/copy VBs by just} updating the \texttt{VBUID} of the corresponding CVT entry with the \texttt{VBUID} of the new VB.}

\subsubsection{Operation of a Memory Load}
\label{sec:load}

Figure~\ref{fig:vbi-uarch} shows the execution of the memory load instruction triggered by the code \texttt{y~=~(*x)}, where the pointer \texttt{x} contains the virtual address consisting of (1)~the index of the corresponding VB in the \sg{process'} CVT, and (2)~the offset within the VB \sg{(\incircle{4} in Figure~\ref{fig:vbi-uarch})}.
\nas{When performing a load operation, the CPU first checks whether \texttt{index} is within the range of the client's CVT.  Next, the CPU needs to fetch the corresponding CVT entry in order to perform the permissions check. The CPU uses a {per-process} small direct-mapped CVT cache to speed up accesses to the client's recently-accessed CVT entries (\pointer{\secref{sec:cvt}}). Therefore, the CPU \sg{looks up} the corresponding CVT cache entry using \texttt{index} as the key \sg{(\incircle{5})}}, \sg{and checks} if (1)~the client has permission to read from the VB, and (2)~\texttt{offset} is smaller than the size of the VB. If either of these checks fail, the CPU raises an exception. If the access is allowed, the CPU constructs the VBI address {by concatenating the \texttt{VBUID}} stored in the CVT entry with \texttt{offset} \sg{(\incircle{6})}. The processor directly uses the generated VBI address to access the on-chip caches. If the data is present in any of the on-chip caches, it is returned to the CPU, thereby completing the load operation.

\nas{\sys performs address translation in parallel with the cache lookup in order to minimize the address translation overhead on the critical path of the access. Accordingly, when an access misses in \sg{the} L2 cache, \nasi{the processor requests the \mtl to perform the VBI-to-physical address translation}. To this end, \mtl fetches the pointer to the VB's translation structure from the VBI Info Table (VIT) entry associated with the VB. \sys uses a VIT cache to speed up accesses to recently-accessed VIT entries (\incircle{7}). In order to facilitate the VBI-to-physical address translation, \mtl employs a translation lookaside buffer (TLB). On a TLB hit, the memory controller accesses the cache line using the physical address in the corresponding TLB entry \sg{(\incircle{8})}. {On a TLB miss, the \mtl performs the address translation by traversing the VB's translation structure \sg{(\incircle{9})}, and inserts the mapping information in{to} the TLB once the physical address is obtained. Next, the memory controller fetches} the corresponding cache line from main memory and returns it to the processor. The processor inserts the cache line into the on-chip caches using the VBI address, and returns the cache line to the CPU to complete the load. \pointer{\secref{sec:mtl}} describes the operation of the \mtl in detail.}

\subsubsection{Memory Deallocation}
\label{sec:mem-dealloc}

{The program can deallocate the memory allocated inside a VB using \texttt{free} (\secref{sec:dynamic_alloc}).} {When a process terminates, the OS traverses the CVT of the process and} detaches all of the VBs attached to the process using the \detachvb instruction. For each VB whose reference count {(stored as part of VIT entry of the VB; see \pointer{\secref{sec:vit}})} drops to zero, the OS informs \sys that the VB is no longer in use via the \disablevb instruction.

\begin{figure}[h!]\small
  \centering
  \begin{tikzpicture}
    \node (enable) [draw,rounded corners=2pt,inner sep=5pt] {\tt \textcolor{blue}{\disablevb} VBUID};
  \end{tikzpicture}
\end{figure}

In response to the \disablevb instruction, the MTL destroys all state associated with VB \texttt{VBUID}. To avoid stale data in the cache, all of the VB's cache lines are invalidated \emph{before} \mhp{the \texttt{VBUID} is reused for another memory allocation}. \mpt{Because} there are a large number of VBs in each size class, \mpt{it is likely that} the disabled \texttt{VBUID} does not need to be reused immediately, and the cache cleanup can be performed lazily in the background.

\subsection{CVT Cache}
\label{sec:cvt}

For every memory operation, the CPU must check if the operation is permitted by accessing the information in the corresponding CVT entry. To exploit locality in the CVT, \sys uses a per-core \emph{CVT cache} to store recently-accessed entries in the client's CVT. The CVT cache is similar to the TLB in existing processors. However, unlike a TLB that caches virtual-to-physical address mappings of page-sized memory regions, the CVT cache maintains information at the VB granularity, and only for VBs that can be accessed by the program. While programs may typically access hundreds or thousands of pages, our evaluations show that most programs only need a few \emph{tens} of VBs to subsume all their data. With the exception of \emph{GemsFDTD} (which allocates 195 VBs),\footnote{\emph{GemsFDTD} performs computations in the time domain on 3D grids. It involves multiple execution timesteps, each of which allocates new 3D grids to store the computation output. Multiple allocations are also needed during the post-processing Fourier transformation performed in \emph{GemsFDTD}.} all applications use fewer than 48 VBs. Therefore, the processor can achieve a near-100\% hit rate even with a 64-entry \emph{direct-mapped} CVT cache, which is faster and more efficient than the large set-associative TLBs employed by modern processors. 

\subsection{Processor, OS, and Process Interactions}
\label{sec:pop-interaction}

\sys handles basic process lifetime operations similar to  current systems. This section describes in detail how these operations work with \sys.

\paragraph{System Booting.}
When the system is booted, the processor initializes the data structures relevant to \sys (e.g., pointers to VIT tables) with the help of the \sgii{MTL} (discussed in \pointer{\secref{sec:mtl}}). An initial ROM program runs as a privileged client, copies the bootloader code from bootable storage to a newly enabled VB, and jumps to the bootloader's entry point. This process initiates the usual sequence of chain loading until the OS is finally loaded into a VB. The OS reads the parameters of \sys, namely, the number of bits of virtual address, the number and sizes of the virtual block size classes, and the maximum number of memory clients supported by the system, to initialize \mhp{the OS-level} memory management subsystem.

\paragraph{Process Creation.} When a binary is executed, the OS creates a new process by associating it with one of the available client IDs. For each \mpt{section of} the binary (e.g., code, static data), the OS (1)~enables the smallest VB that can fit the contents of the \mpt{section} and associates the VB with the appropriate properties using the \enablevb instruction, (2)~attaches itself to the VB with write permissions using the \attachvb instruction, (3)~copies the contents from the application binary into the VB, and (4)~detaches itself from the VB using the \detachvb instruction. The OS then attaches the client to the newly enabled VBs and jumps to program's entry point.

\paragraph{Shared Libraries.}
The OS loads the executable code of each shared library into a separate VB. While \mpt{a} shared library can dynamically allocate data using the \requestvas system call, any \mpt{static per-process} data associated with the library should be loaded \mhp{into} a separate VB for each process that uses the library. In existing systems, access to \mpt{static} data is \mpt{typically} performed using PC-relative addressing. \sys~\mpt{provides an analogous memory} addressing mode that we call \emph{CVT-relative addressing}. In this addressing mode, the CVT index of a memory reference is specified relative to the CVT index of the VB containing the reference. Specifically, in shared libraries, all references to static data use +1 CVT-relative addressing, i.e., the CVT index of the data is one more than the CVT index of the code. After process creation, the OS iterates over the list of shared libraries requested by the process. For each shared library, the OS attaches the client to the VB containing the corresponding library code and ensures that the subsequent CVT entry is allocated to the VB containing the static data associated with the shared library. This solution avoids the \mpt{need to perform load-time relocation for each data reference in the executable code}, although VBI can use relocations in the same manner as current systems, \mpt{if required}.

\paragraph{Process Destruction.}
When a process terminates, the OS deallocates all VBs for the process using the mechanism described in \pointer{\secref{sec:mem-dealloc}}, and then frees the client ID for reuse.

\paragraph{Process Forking.}
When a process forks, all of its memory state must be replicated for the newly created process. In \sys, forking entails creating copies of all the private VBs attached to a process. To reduce the overhead of this operation, \sys introduces the following instruction:
\begin{figure}[h!]\small
  \centering
  \begin{tikzpicture}
    \node [draw,rounded corners=2pt,inner sep=5pt] {\tt \textcolor{blue}{\clonevb} SVBUID, DVBUID};
  \end{tikzpicture}
\end{figure}

\clonevb  instructs \sys to make the destination VB \texttt{DVBUID} a clone of the source VB \texttt{SVBUID}. To efficiently implement \clonevb, the MTL marks all translation structures and physical pages of the VB as copy-on-write, and lazily copies the relevant regions if they receive a write operation.\footnote{{The actual physical copy can be accelerated using in-DRAM copy mechanisms such as RowClone~\cite{seshadri2013},  LISA~\cite{chang16}, {and NoM~\cite{nom2020}}.}}

When forking a process, the OS first copies all CVT entries of the parent to the CVT of the child so that the child VBs have the same CVT index
as the parent VBs. This maintains the validity of the pointers
in the child VBs after cloning. Next, for each CVT entry corresponding to a private VB (shared VBs are already enabled), the OS (1)~enables a new VB of the same size class and executes the \clonevb instruction, and (2)~updates the \texttt{VBUID} in the CVT entry to point to the newly enabled clone. The fork returns after all the \clonevb operations are completed.

\paragraph{VB Promotion.}
\label{sec:vb-promotion}
As described in \pointer{\secref{sec:dynamic_alloc}}, when a program runs out of memory for a data structure within the assigned VB, the OS can automatically promote the data structure to a VB of higher size class. To perform such a promotion, the OS first suspends the program. It enables a new VB of the higher size class, and executes the \promotevb instruction.
\begin{figure}[h!]\small
  \centering
  \begin{tikzpicture}
    \node [draw,rounded corners=2pt,inner sep=5pt] {\tt \textcolor{blue}{\promotevb} SVBUID, LVBUID};
  \end{tikzpicture}
\end{figure}

In response to this instruction, \sys first flushes all dirty cache lines from the smaller VB with the unique ID of \texttt{SVBUID}. This operation can be sped up using structures like the Dirty Block Index~\cite{dbi}. \sys then copies all the translation information from the smaller VB appropriately to the larger VB with the unique ID of \texttt{LVBUID}. After this operation, in effect, the early portion of the larger VB is mapped to the same region in the physical memory as the smaller VB. The remaining portions of the larger VB are unallocated and can be used by the program to expand its data structures {and allocate more memory using \texttt{malloc}}. \sys updates the entry in the program's CVT {that points} to \texttt{SVBUID} to now point to \texttt{LVBUID}.

\subsection{\mtlfull}
\label{sec:mtl}

The \mtlfull (\mtl) centers around the VB Info Tables (VITs){, which store the metadata associated with each VB}. In this section, we discuss (1)~the design of the VITs, (2)~the two main responsibilities of the \mtl; memory allocation and address translation, and (3) the hardware complexity of the \mtl.

\subsubsection{\vit (\vitshort)}
\label{sec:vit}

{As \pointer{\secref{sec:vb-detail}} briefly describes, MTL uses a set of VB Info Tables (VITs) to maintain information about VBs. Specifically, 
for each VB {in the system}, a \vit stores an entry that consists of (1)~an \emph{enable} bit, which indicates if the VB is currently assigned to a process; (2)~\texttt{props}, a bitvector that describes the VB properties; \nas{(3)~the number of processes attached to the VB (i.e., a reference count);} (4)~the type of VBI-to-physical address translation structure being used for the VB; and (5)~a pointer to the translation structure.} For ease of access, the MTL maintains a \emph{separate} VIT for each size class. The ID of a VB within its size class (VBID) is used as an index into the corresponding VIT. When a VB is enabled (using \enablevb), the MTL {finds the corresponding VIT and entry using the \texttt{SizeID} and \texttt{VBID}, respectively (both extracted from \texttt{VBUID}). \mtl then sets the} \emph{enabled} bit of the entry and updates \texttt{props}. \nas{The reference counter of the VB is also set to 0, indicating that no process is attached to this VB.} The type and pointer of the translation structure {of the VB} are updated {in its VIT entry} at the time of physical memory allocation (as we discuss in \pointer{\secref{sec:flexible-translations}}). {Since a VIT contains {entries for the} VBs of only a single size class, the number of entries in each VIT equals the number of VBs that the associated size class supports (\pointer{\secref{sec:vb-detail}}). However, }\sys limits the size of each \vit by storing entries only \mhp{up to the currently-enabled} VB with the largest \texttt{VBID} {in the size class associated with that \vit}. The OS ensures that the table does not become prohibitively large by reusing previously-disabled VBs for subsequent requests (\pointer{\secref{sec:mem-dealloc}}).

\subsubsection{{Base Memory Allocation and Address Translation}}
\label{sec:base}
{Our \emph{base} memory allocation algorithm} allocates physical memory at 4~KB granularity. Similar to \xeightsix~\cite{intelx86manual}, {Our \emph{base} address translation mechanism} stores VBI-to-physical address translation information in multi-level tables. However, unlike the 4-level page tables in \xeightsix, \sys uses tables with varying number of levels according to the size of the VB. {For example, a 4~KB VB does not require a translation structure (i.e., can be direct-mapped) since 4~KB is the minimum granularity of meomry allocation. On the other hand, a 128~KB VB requires a one-level table for translating address to 4~KB regions}. As a result, smaller VBs require fewer memory accesses to serve a TLB miss. For each VB, the \vitshort stores a pointer to the address of the root of the multi-level table (or the base physical address of the directly mapped VBs). 

\subsubsection{\mhp{\mtl} Hardware Complexity}

We envision the \mtl as software running on a programmable low-power core within the memory controller. While conventional OSes are responsible for memory allocation, virtual-to-physical mapping, and memory protection, the \mtl does not need to deal with protection, so we expect the \mtl code to be simpler than typical OS memory management software. As a result, the complexity of the \mtl hardware is similar to that of prior proposals such as Pinnacle~\cite{pinnacle} (commercially available) and Page Overlays~\cite{page_overlays}, which perform memory allocation and remapping in the memory controller. While both Pinnacle and Page Overlays are hardware solutions, VBI provides flexibility by making the \mtl programmable, thereby allowing software updates for different memory management policies (e.g., address translation, mapping, migration, scheduling). Our goal in this work is to understand the potential of hardware-based memory allocation and address translation.

\section{Allocation and Translation Optimizations}
\label{sec:mtl_optimizations}

The \mtl employs three techniques to optimize the base memory allocation and address translation {described in \pointer{\secref{sec:base}}}. \nasi{We explain these techniques in the following subsections.}

\subsection{Delayed Physical Memory Allocation}
\label{sec:delayed-allocation}

As described in \pointer{\secref{sec:optimizations}}, \sys delays physical memory allocation for a VB (or a region of a VB) until a dirty cache line from that {VB (or a region of the VB)} is evicted from the last-level cache (LLC). This optimization is enabled by the fact that \sys uses VBI address directly to access \mhp{\emph{all}} on-chip caches. Therefore, a cache line does \emph{not} need to be backed by a physical memory mapping in order to be accessed.

{In this approach,} when a VB is enabled, \sys does not immediately allocate physical memory to the VB. On an LLC miss {to the VB}, \sys checks the status of the VB in its corresponding VIT entry. If there is no physical memory backing the data, \sys does one of two things. (1)~If the VB corresponds to a memory-mapped file or if the required data was {allocated before but }swapped out to a backing store, then \sys allocates physical memory for the region, interrupts the OS to copy the relevant data from storage into the allocated memory, and then returns the relevant cache line to the processor. (2)~If this is the first time the cache line is being accessed from memory, \sys simply returns a zeroed cache line without allocating physical memory to the VB.

{On a dirty cache line {writeback from the LLC}, if physical memory is yet to be allocated for the region that the cache line maps to, \sys first allocates physical memory for the region, and then performs the writeback. \sys allocates only the region of the VB containing the evicted cache line. As \pointer{\secref{sec:base}} describes, our base memory allocation mechanism allocates physical memory at a 4~KB granularity. Therefore, the region allocated for the evicted cache line is 4~KB.} \pointer{\secref{sec:early-reservation}} describes an optimization that eagerly \emph{reserves} {a larger amount of} physical memory for a VB {during allocation}, to reduce {the} overall translation overhead.

\subsection{Flexible \nasi{Address} Translation Structures}
\label{sec:flexible-translations}
For each VB, \sys chooses one of three types of address translation structures, depending on the needs of the VB and the physical memory availability. The first type \emph{directly} maps the VB to physical memory when enough contiguous memory is available. With this mapping, a single TLB entry is sufficient to maintain the translation for the entire VB. The second type uses a single-level table, where the VB is divided into equal-sized blocks of one of the supported size classes. Each entry in the table maintains the mapping for the corresponding block. This mapping exploits the fact that \mhp{a} majority of the data structures are densely allocated inside their respective VBs. With \mhp{a} single-level table, the mapping for any region of the VB can be retrieved with a single memory access. The third type, suitable for sparsely-allocated VBs, is {our base address translation mechanism (described in \pointer{\secref{sec:mtl}}), which uses multi-level page tables where the table depth is chosen based on the size of the VB.}

{\sgii{In our evaluation, we implement a flexible mechanism that} statically chooses a translation structure type based on the size of the VB. Each 4~KB VB is directly mapped. 128~KB and 4~MB VBs use a single-level table. VBs of a larger size class use a multi-level table with as many levels as necessary to map the VB using 4~KB pages.\footnote{{For fair comparison with conventional virtual memory, our evaluations use a 4~KB granularity to map VBs to physical memory. However, \sys can flexibly map VBs at the granularity of any available size class.}} The {\emph{early reservation}} optimization (described in \pointer{\secref{sec:early-reservation}}) improves upon this static policy by dynamically choosing a translation structure type from the three types mentioned above based on the available contiguous physical memory. While we evaluate table-based translation structures in this work, \sys can be easily extended to support other structures (e.g., customized per-application translation structures as proposed in DVMT~\cite{vm19}).}

Similar to \xeightsix, \sys uses multiple types of TLBs to cache mappings of different granularity. The type of translation structure used for a VB is stored in the \vitshort and is cached in the on-chip \vitshort Cache. This information enables \sys to access the right type of TLB. For a fair comparison, our evaluations use the same TLB type and size for all baselines and variants of \sys.

\subsection{Early Reservation of Physical Memory}
\label{sec:early-reservation}

\sys can perform early reservation of the physical memory for a VB. {To this end, \sys reserves (but does not allocate) physical memory for the entire VB at the time of memory allocation, and treats the VB as \emph{directly mapped} by serving future memory allocation requests for that VB from that contiguous reserved region.} This optimization is inspired by prior work on super-page management~\cite{reserve}, which reserves a larger contiguous region of memory than the requested size, and upgrades the allocated pages to larger super-pages when enough contiguous pages are allocated in that region.

For \sys's early reservation optimization, at the time of the \emph{first} physical memory allocation request for a VB, the \mtl checks if there is enough contiguous free space in physical memory to fit the entire VB. If so, it allocates the requested memory from that contiguous space, and marks the remaining free blocks in that contiguous space as reserved for that specific VB. In order to reduce internal fragmentation when free physical memory is running low, physical blocks reserved for a VB may \mhp{be} used by \mhp{another} VB when no unreserved blocks are available.  As a result, the \mtl uses a three-level priority when allocating physical blocks: (1)~free blocks reserved for the VB that is demanding allocation, (2)~unreserved free blocks, and (3)~free blocks reserved for other VBs. A VB is considered directly mapped as long as all its allocated memory is mapped to a single contiguous region of memory, thereby requiring just a single TLB entry for the entire VB. If there is not enough contiguous physical memory available to fit the entire VB, the early reservation mechanism allocates the VB sparsely by reserving blocks of the largest size class that can be allocated contiguously.

{With the early reservation approach, memory allocation is performed at a different granularity than mapping, which enables \sys to benefit from larger mapping granularities and \mhp{thereby} minimize the address translation latency, while eliminating memory allocation for regions that may never be accessed. }
To support the early reservation mechanism, \sys uses the Buddy algorithm~\cite{buddy,knowlton65} to manage free and reserved \nasi{regions} of different {size classes}.

\section{VBI in Other System Architectures}

\sys is designed to easily and efficiently function in various system designs. \mhp{We} describe the implementation of \sys in two important examples of modern system architectures: virtualized environments and multi-node systems.

\subsection{Supporting Virtual Machines}
\label{sec:virtual-machines}

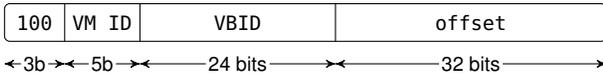
\begin{figure}[h]
  \centering
  \begin{tikzpicture}[inner sep=1pt,outer sep=0pt,>=stealth', font=\sffamily]

\draw [rounded corners=2pt] (0,0) rectangle (8,0.5);
\draw (4.4,0) -- ++(0,0.5);
\draw (0.8,0) -- ++(0,0.5);
\draw (1.8,0) -- ++(0,0.5);

\node at (0.4,0.25) {\small\texttt{100}};
\node at (1.3,0.25) {\small{\texttt{VM ID}}};
\node at (3.1,0.25) {\small\texttt{VBID}};
\node at (6.2,0.25) {\small\texttt{offset}};

\node (s) at (0.4,-0.3) {\small{3b}};
\node (m) at (1.3,-0.3) {\small{5b}};
\node (v) at (3.1,-0.3) {\small{24 bits}};
\node (o) at (6.2,-0.3) {\small{32 bits}};

\draw [->] (s.west) -- (0,-0.3);
\draw [->] (s.east) -- (0.8,-0.3);
\draw [->] (m.west) -- (0.8,-0.3);
\draw [->] (m.east) -- (1.8,-0.3);
\draw [->] (v.west) -- (1.8,-0.3);
\draw [->] (v.east) -- (4.4,-0.3);
\draw [->] (o.west) -- (4.4,-0.3);
\draw [->] (o.east) -- (8,-0.3);

\end{tikzpicture}
  \caption{Partitioning the VBI address space among virtual machines, using the 4~GB size class (\texttt{100}) as an example.}
  \label{fig:vaddr-vm}
  \vspace{-5pt}
\end{figure}

\sys implements address space isolation between virtual machines (VMs) by \emph{partitioning} the global VBI address space among multiple VMs and the host OS. To this end, \sys reserves a few bits in the VBI address for the \texttt{VM ID}. Figure~\ref{fig:vaddr-vm} shows how \sys implements this for a system supporting 31 virtual machines (ID 0 is reserved for the host). In the {VBI} address, the 5 bits following the size class bits are used to denote the \texttt{VM ID}. For every new virtual machine in the system, the host OS assigns a \texttt{VM ID} to be used by the guest OS while assigning virtual blocks to processes inside the virtual machine. \sys partitions client IDs using a similar approach. With address space division between VMs, a guest VM is unaware that it is virtualized, and it can allocate/deallocate/access VBs {\emph{without} having to} coordinate with the host OS. Sharing VBs across multiple VMs is possible, but requires explicit coordination with the host OS.

\subsection{Supporting Multi-Node Systems}
\label{sec:multi-node}

There are many ways to implement \sys in multi-node systems. Our initial approach provides each node with its own \mtl. \sys equally partitions VBs of each size class among the {\mtl}s, with the higher order bits of \texttt{VBID} indicating the \sgii{VB's} home \mtl. The home \mtl of a VB is the only \mtl that manages the VB's physical memory allocation and address translation. When allocating a VB to a process, the OS attempts to ensure that the \sgii{VB's} home \mtl is in the same node as the core executing the process. During phase changes, the OS can seamlessly migrate data from a VB hosted by one \mtl to a VB hosted by another \mtl. We leave the evaluation of this approach and exploration of other ways of integrating \sys with multi-node systems to future work.

\section{Evaluation}
\label{sec:methodology2}

We evaluate VBI for two concrete use cases. First, we evaluate how \sys reduces address translation overheads in native and virtualized environments (\pointer{\secref{sec:4k-perf}} and \pointer{\secref{sec:2m-perf}}\mhp{, respectively}). Second, we evaluate the benefits that \sys offers in harnessing the full potential of two \mhp{main memory} architectures that are tightly dependent on the \nas{data mapping}: (1)~a hybrid PCM--DRAM memory architecture; and (2)~TL-DRAM\mhp{\cite{tldram}}, a \mhp{heterogeneous-latency} DRAM \onurii{architecture} (\pointer{\secref{sec:tld}}). 

\subsection{Methodology}
\label{sec:methodology}

For our evaluations, we use a heavily-customized version of Ramulator~\cite{ramulator} to faithfully model all components of the memory subsystem (including TLBs, page tables, the page table walker, and the page walk cache), as well as the functionality of memory management calls (e.g., \texttt{malloc}, \texttt{realloc}, \texttt{free}). We have released this modified version of Ramulator~\cite{vbi}. Table~\ref{tbl:simconfig} summarizes the main simulation parameters. Our workloads consist of benchmarks from SPECspeed 2017~\cite{spec2017}, SPEC CPU 2006~\cite{spec2006}, TailBench~\cite{tailbench}, and Graph 500~\cite{graph500}. We identify representative code regions for the SPEC benchmarks using SimPoint~\cite{sim-point}. For TailBench applications, we skip the first five billion instructions. For Graph 500, we mark the region of interest directly in the source code. We use an Intel Pintool~\cite{luk2005} to collect traces of the representative regions of each of our benchmarks. For our evaluations, we first warm up the system with 100~million instructions, and then run the benchmark for 1~billion instructions.

\begin{table}[h]\scriptsize
  \centering
  \vspace{3pt}
  \setlength{\aboverulesep}{0pt}
    \setlength{\belowrulesep}{0pt}
  \def\arraystretch{0.9}
   \begin{tabular}{ll} 
        \toprule
        \textbf{CPU} & 4-wide issue, OOO, 128-entry ROB\\
        \cmidrule(rl){1-2}
        \textbf{L1 Cache} & 32~KB, 8-way associative, 4 cycles \\
        \cmidrule(rl){1-2}
        \textbf{L2 Cache} & 256~KB, 8-way associative, 8 cycles\\
        \cmidrule(rl){1-2}
        \textbf{L3 Cache} & 8~MB (2~MB per-core), 16-way associative, 31 cycles\\
        \cmidrule(rl){1-2}
        \multirow{2}{*}{\textbf{L1 DTLB}} & 4~KB pages: 64-entry, fully associative\\
        & 2~MB pages: 32-entry, fully associative\\
        \cmidrule(rl){1-2}
        \textbf{L2 DTLB} & 4~KB and 2~MB pages: 512-entry, 4-way associative\\
        \cmidrule(rl){1-2}
        \textbf{Page Walk Cache} & 32-entry, fully associative\\
        \cmidrule(rl){1-2}
        \multirow{2}{*}{\textbf{DRAM}} & DDR3-1600, 1 channel, 1 rank/channel\\
         & 8 banks/rank, open-page policy\\
        \cmidrule(rl){1-2}
        \multirow{1}{*}{\textbf{DRAM Timing~\cite{dramtiming}}} & tRCD=5cy, tRP=5cy, tRRDact=3cy, tRRDpre=3cy\\
        \cmidrule(rl){1-2}
        \textbf{PCM} & PCM-800, 1 channel, 1 rank/channel, 8 banks/rank\\
        \cmidrule(rl){1-2}
        \multirow{1}{*}{\textbf{PCM Timing~\cite{lee2009}}} & tRCD=22cy, tRP=60cy, tRRDact=2cy, tRRDpre=11cy\\
        \bottomrule
    \end{tabular}

  \caption{Simulation configuration.}
  \label{tbl:simconfig}
  \vspace{-10pt}
\end{table}%

\begin{figure*}[h]
  \centering
  \includegraphics[scale=0.75]{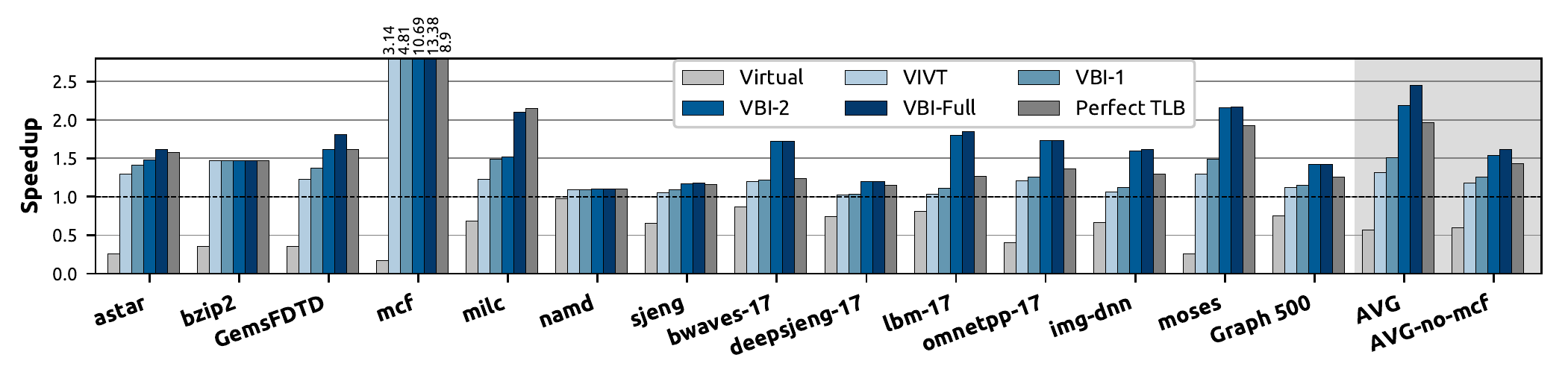}
  \vspace{-10pt}
  \caption{Performance of systems with 4KB pages (normalized to \textsf{Native}).}
  \vspace{0pt}
  \label{fig:4k-perf}
\end{figure*}

\subsection{Use Case 1: Address Translation}
\label{sec:vm-perf}

We evaluate the performance of seven baseline systems to compare with \sys:
(1)~\textbf{\sffamily Native}: applications run natively on an \xeightsix system with only 4~KB pages;
(2)~\textbf{\sffamily Native-2M}: \textsf{Native} but with only 2~MB pages; 
(3)~\textbf{\sffamily Virtual}: applications run inside a virtual machine with only 4~KB pages;
(4)~\textbf{\sffamily Virtual-2M}: \textsf{Virtual} but with only 2~MB pages;\footnote{We augment this system with a 2D page walk cache, which is shown to improve the performance of guest workloads~\cite{vm11}.} 
(5)~\textbf{\sffamily Perfect TLB}: an unrealistic version of \textsf{Native} with no L1 TLB misses \onurii{(i.e., no address translation overhead)};
(6)~\textbf{\sffamily VIVT}: \textsf{Native} with VIVT on-chip caches; and
(7)~\onurii{\textbf{\sffamily Enigma-HW-2M}}: applications run natively in a system with Enigma~\cite{enigma}. \onurii{Enigma uses a system-wide unique intermediate address space to defer address translation until data must be retrieved from physical memory. A centralized translation cache (CTC) at the memory controller \sgii{performs intermediate-to-physical address translation}. However, \sgii{unlike \sys, Enigma asks the OS to perform the translation on a CTC miss, and} to explicitly manage address mapping. Therefore, Enigma's benefits do not seamlessly extend to programs running inside a virtual machine. We evaluate Enigma with a 16K-entry centralized translation cache (CTC) that we enhance with hardware-managed page walks and 2 MB pages.}

We evaluate the performance of three \sys systems:
(1)~\textbf{\sffamily \mbox{\sys-1}}: \onurii{inherently virtual caches (\pointer{\secref{sec:optimizations}}) along with} \onurii{our} \onurii{\emph{flexible translation mechanism} \onurii{that} maps VBs using \sgii{a 4~KB granularity}~(\pointer{\secref{sec:base}})} ,
(2)~\textbf{\sffamily \mbox{\sys-2}}: \textsf{\sys-1} with \emph{delayed physical memory allocation} \onurii{(allocates the 4~KB region of the VB that the dirty cache line evicted from the last-level cache belongs to)}. (\pointer{\secref{sec:delayed-allocation}}), and
(3)~\textbf{\sffamily \mbox{\sys-Full}}: \textsf{\sys-2} with \emph{early reservation} (\pointer{\secref{sec:early-reservation}}). \textsf{\sys-1} and \textsf{\sys-2} manage memory at 4~KB granularity, while \textsf{\sys-Full} uses early reservation to support all of the size classes listed in \pointer{\secref{sec:vb-detail}} for VB allocation, providing similar benefits to large page support \onurii{and direct mapping}. \onurv{We first present results comparing \textsf{\sys-1} and \textsf{\sys-2}} with \textsf{Native}, \textsf{Virtual}, \textsf{VIVT}, and \textsf{Perfect TLB} (\pointer{\secref{sec:4k-perf}}). We then present results comparing \textsf{\sys-Full} with \textsf{Native-2M}, \onurii{\textsf{Enigma-HW-2M}}, and \textsf{Perfect TLB} (\pointer{\secref{sec:2m-perf}}).

\subsubsection{Results with 4~KB Pages}
\label{sec:4k-perf}

Figure~\ref{fig:4k-perf} plots the performance of \textsf{Virtual}, \textsf{VIVT}, \textsf{\sys-1}, \textsf{\sys-2}, and \textsf{Perfect TLB} normalized to the performance of \textsf{Native}, for a single-core system. \onurii{We also show \textsf{\sys-Full} as a reference} \onurv{that shows the full potentials of \sys which \textsf{\sys-1} and \textsf{\sys-2} do not enable}. \emph{mcf} has an overwhelmingly high number of TLB misses. Consequently, mechanisms that reduce TLB misses greatly improve \emph{mcf}'s performance, to the point of skewing the \onurvii{average} significantly. Therefore, the figure also presents the \onurvii{average} speedup without \emph{mcf}. We draw five observations from the figure.

First, \textsf{\sys-1} outperforms \textsf{Native} by 50\%, averaged across all benchmarks (25\% without \emph{mcf}). This performance gain is a direct result of \onurii{(1) inherently virtual on-chip caches in \sys that reduce the number of address translation requests, and (2) }fewer levels of address translation for smaller VBs, \onurv{which reduces} the number of translation-related memory accesses (i.e., page walks).

Second, \textsf{Perfect TLB} serves as an upper bound for the performance benefits of \textsf{\sys-1}. \onurii{However, by employing flexible translation structures,} \textsf{\sys-1} bridges the performance gap between \textsf{Native} and \textsf{Perfect TLB} by 52\%\mhp{, on average}.

Third, when accessing regions for which no physical memory is allocated yet, \textsf{\sys-2} avoids \emph{both} the memory \onurii{requests themselves and any} translation-related memory accesses \onurvii{for those requests}. \onurii{Therefore, \textsf{\sys-2} enables benefits over and beyond solely reducing the number of page walks, as it further improves the overall performance by reducing the number of memory requests accessing the main memory as well.} Consequently, for many memory-intensive applications, \textsf{\sys-2} outperforms \textsf{Perfect TLB}. Compared to \textsf{Perfect TLB}, \textsf{\sys-2} reduces the total number of DRAM accesses (including the translation-related memory accesses) by 62\%, averaged across applications that outperform \textsf{Perfect TLB}, and by 46\% across all applications. Overall, \textsf{\sys-2} outperforms \textsf{Native} by an average of 118\% (53\% without \emph{mcf}).

Fourth, by performing address translations \emph{only for and in parallel with} LLC accesses, \textsf{VIVT} outperforms \textsf{Native} by 31\% on average (17\% without \emph{mcf}). This performance gain is due to reducing the number of translation requests and therefore decreasing the number of TLB misses using VIVT caches. However, \textsf{\sys-1} and \textsf{\sys-2} gain an extra 19\% and 87\% performance on average, respectively, over \textsf{VIVT}. These improvements highlight \sys's ability to improve performance beyond \mhp{only} employing VIVT caches.

Finally, our results indicate \onurii{that} due to considerably higher translation overhead, \textsf{Virtual} significantly slows down applications compared to \textsf{Native} (44\% on average). As described in \pointer{\secref{sec:optimizations}}, once an application running inside a virtual machine \onurvi{is attached to its VBs}, \sys incurs no additional translation overhead compared to running natively. As a result, in virtualized environments that use only 4K pages, \textsf{\sys-1} and \textsf{\sys-2} achieve an average performance of 2.6$\times$ and 3.8$\times$, respectively, compared to \textsf{Virtual}.

\onurthird{We conclude that even when mapping and allocating VBs using 4~KB granularity only,} both \textsf{\sys-1} and \textsf{\sys-2} provide \mhp{large} benefits over a wide range of baseline systems, due to their effective optimizations to reduce address translation and memory allocation overheads. 
\onurvi{\textsf{\sys-Full} further improves performance by mapping VBs using larger granularities (as we elaborate in \secref{sec:2m-perf}).}

\subsubsection{Results with Large Pages}
\label{sec:2m-perf}

\begin{figure}[h]
  \centering
  \vspace{-3pt}
  \includegraphics[width=\linewidth]{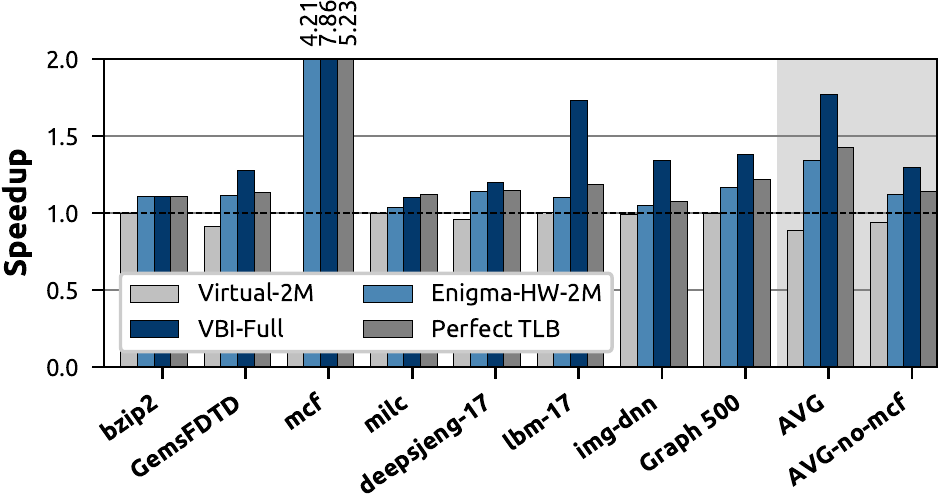}
  \vspace{-16pt}
  \caption{Performance with large pages (\mhp{norm.} to \textsf{Native-2M}).}
  \label{fig:2m-perf}
  \vspace{-4pt}
\end{figure}

Figure~\ref{fig:2m-perf} plots the performance of \textsf{Virtual-2M}, \onurii{\textsf{Enigma-HW-2M}}, \textsf{VBI-Full}, and \textsf{Perfect TLB} normalized to the performance of \textsf{Native-2M}. We enhance the original design of Enigma~\cite{enigma} by replacing the OS system call handler for address translation on a CTC miss with a completely \mhp{hardware-managed} address translation, similar to \sys. For legibility, the figure shows results for only a subset of the applications. However, the chosen applications capture the behavior of all the applications, and the \onurvii{average} (and \onurvii{average} without \emph{mcf}) is calculated across all evaluated applications. We draw three observations from the figure. 

\onurvi{First, managing memory at 2~MB granularity improves the performance of applications compared to managing memory at 4~KB granularity.  This is} because (1)~the larger page size lowers the average TLB miss count \onurvi{(e.g., 66\% lower for \textsf{Native-2M} compared to \textsf{Native}), and (2)~requires fewer page table accesses on average to serve TLB misses (e.g., 73\% fewer for \textsf{Native-2M} compared to \textsf{Native})}.

\onurvi{Second}, \onurii{\textsf{Enigma-HW-2M}} improves overall performance for programs running \emph{natively} on the system by 34\% compared to \textsf{Native-2M}, averaged across all benchmarks (including mcf). The performance gain is a direct result of
(1)~the very large CTC (16K~entries), which reduces the number of translation-related memory accesses by 89\% on average compared to \textsf {Native-2M}; and 
(2)~our hardware-managed address translation enhancement, which removes the costly system calls on each page walk request.

Third, \textsf{\sys-Full}, with all three of our optimizations in \pointer{\secref{sec:mtl_optimizations}}, maps most VBs using direct mapping, thereby significantly reducing the number of TLB misses and translation-related memory accesses compared to \textsf{Native-2M} (on average by 79\% and 99\%, respectively). In addition, \textsf{\sys-Full} retains the benefits of \textsf{\sys-2}, which reduces the number of \emph{overall} DRAM accesses. \textsf{\sys-Full} reduces the \emph{total} number of DRAM accesses (including translation-related memory accesses) by 56\% on average compared to \textsf{Perfect TLB}. Consequently, \textsf{\sys-Full} outperforms all four comparison points including \textsf{Perfect TLB}. Specifically, \textsf{\sys-Full} improves performance by 77\% compared to \textsf{Native-2M}, 43\% compared to \onurii{\textsf{Enigma-HW-2M}} and 89\% compared to \textsf{Virtual-2M}.

We conclude that \onurii{by employing all of the optimizations that it enables, \sys} significantly outperforms all of our baselines in both native and virtualized environments.

\subsubsection{Multicore Evaluation}
\begin{figure}[t]
  \centering
  \vspace{0pt}
  \includegraphics[width=0.93\linewidth]{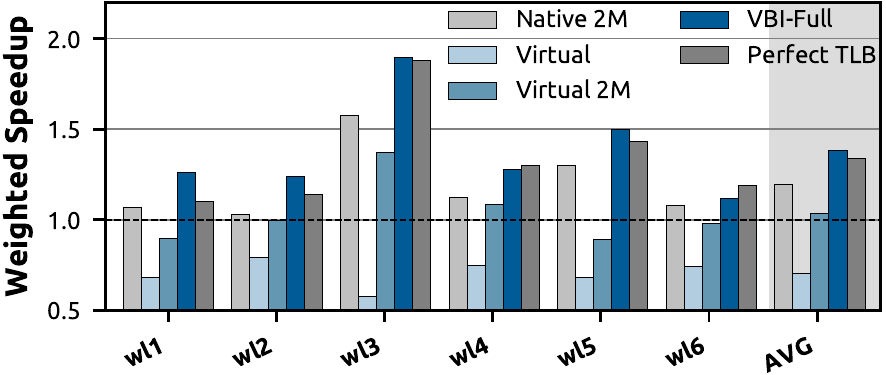}
  \vspace{-5pt}
  \caption{Multiprogrammed workload performance (normalized to \textsf{Native}).}
  \label{fig:mp}
  \vspace{-3pt}
\end{figure}

Figure~\ref{fig:mp} compares the weighted speedup of \textsf{VBI-Full} against four baselines in a quad-core system. We examine six different workload bundles, listed in Table~\ref{tbl:wl}, which consist of the applications studied in our single-core evaluations.  From the figure, we make two observations.  First, averaged across all bundles, \textsf{VBI-Full} improves performance by 38\% and 18\%, compared to \textsf{Native} and \textsf{Native-2M}, respectively. Second, \textsf{VBI-Full} outperforms \textsf{Virtual} and \textsf{Virtual-2M} by an average 67\% and 34\%, respectively. We conclude that the benefits of \sys persist even in the presence of higher memory load in multicore systems.

\begin{table}[b]
\scriptsize
\setlength\tabcolsep{3pt}
  \centering
    \setlength{\aboverulesep}{0pt}
    \setlength{\belowrulesep}{0pt}
  \begin{tabular}{ll|ll}
  \toprule
  \mhp{{\textbf wl1}} &  deepsjeng, omnetpp, bwaves, lbm & \mhp{{\textbf wl4}} &  milc, namd, GemsFDTD, bzip2\\
  \mhp{{\textbf wl2}} &  graph500, astar, img-dnn, moses & \mhp{{\textbf wl5}} &  bzip2, GemsFDTD, sjeng, mcf\\
  \mhp{{\textbf wl3}} &  mcf, GemsFDTD, astar, milc & \mhp{{\textbf wl6}} &  namd, bzip2, astar, sjeng\\
  \bottomrule
\end{tabular}
  \caption{Multiprogrammed workload bundles.}
  \label{tbl:wl}
\end{table}
  \vspace{0pt}

\subsection{Use Case 2: Memory Heterogeneity}
\label{sec:tld}

As mentioned in \pointer{\secref{sec:intro}}, extracting the best performance from heterogeneous-latency DRAM \onurii{architectures}~\cite{charm, diva-dram, tldram, dynsub, chang16, chang2016low, kim2018solar,clrdram,seshadri2013} and hybrid memory architectures~\cite{yoon2012, refree, raoux2008, li2017utility, dhiman2009pdram, ramos11, het2, zhang2009exploring,chop,banshee} critically depends on \onurii{mapping data to the memory that suits the data requirements, and migrating data as its} \onurvii{requirements} change. We quantitatively show the performance benefits of \sys in exploiting heterogeneity by evaluating (1)~a PCM--DRAM hybrid memory~\cite{ramos11}; and (2)~TL-DRAM~\cite{tldram}, a \mhp{heterogeneous-latency} DRAM \onurii{architecture}. We evaluate five systems:
(1)~\textsf{\sys PCM--DRAM} and (2)~\textsf{\sys TL-DRAM}, in which \sys maps and migrates frequently-accessed \onurthird{VBs} to the low-latency memory (the fast memory region in the case of TL-DRAM);
(3)~\textsf{Hotness-Unaware PCM--DRAM} and (4)~\textsf{Hotness-Unaware TL-DRAM}, where the mapping mechanism is unaware of the hotness (i.e., the access frequency) of the data and therefore do not necessarily map the frequently-accessed data to the fast region; and
(5)~\textsf{IDEAL} in each plot refers to an unrealistic perfect mapping mechanism, which uses oracle knowledge to always map \mhp{frequently-accessed} data to the \mhp{fast portion of} memory.

\begin{figure}[h]
  \centering
  \includegraphics[width=\linewidth]{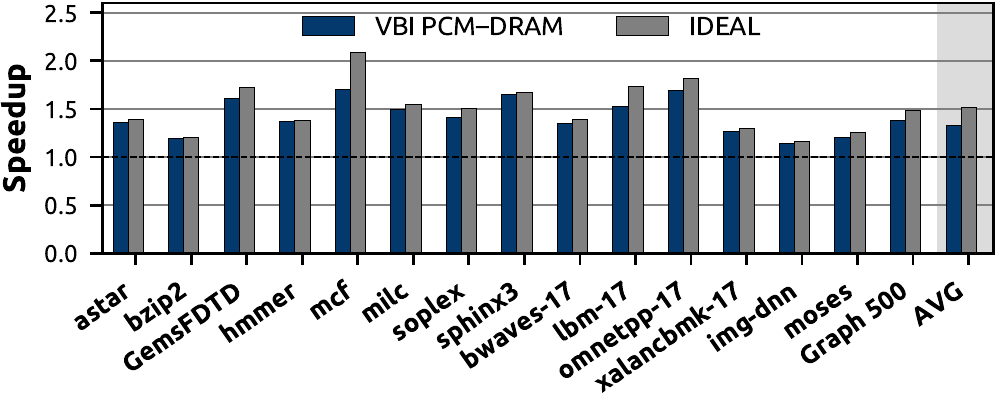}
  \vspace{-16pt}
  \caption{Performance of \sys PCM-DRAM (normalized to \mhp{data-hotness-unaware} mapping).}
  \vspace{-2pt}
  \label{fig:pcm}
\end{figure}

\begin{figure}[h]
  \centering
  \includegraphics[width=\linewidth]{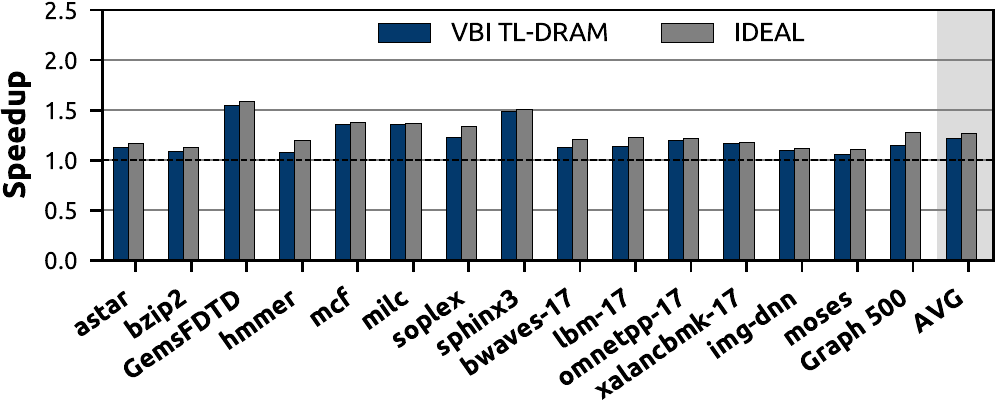}  
  \vspace{-13pt}
  \caption{Performance of \sys TL-DRAM (normalized to \mhp{data-hotness-unaware} mapping).}
  \label{fig:tldram}
  \vspace{-6pt}
\end{figure}

Figures~\ref{fig:pcm} and~\ref{fig:tldram} show the speedup obtained by \sys-enabled mapping over the hotness-unaware mapping in a PCM--DRAM hybrid memory and in TL-DRAM, respectively. We draw three observations from the figures. First, for PCM--DRAM, \textsf{\sys PCM--DRAM} improves performance by 33\% on average compared to the \textsf{Hotness-Unaware PCM--DRAM}, by accurately mapping the frequently-accessed data structures to the low-latency DRAM. Second, by mapping \mhp{frequently-accessed} data to the \mhp{fast DRAM regions}, \textsf{\sys TL-DRAM} takes \onurii{better} advantage of the benefits of TL-DRAM, with a performance improvement of 21\% on average compared to \textsf{Hotness-Unaware TL-DRAM}. Third, \textsf{\sys TL-DRAM} performs only 5.3\% slower than \textsf{IDEAL}, which is the upper bound of performance achieved by mapping hot data to the fast \mhp{regions} of DRAM. 

We conclude that \sys is effective for enabling efficient data mapping and migration in heterogeneous memory systems.
\section{Related Work}
\label{sec:related}

\onurthird{To our knowledge, \sys is the first virtual memory framework to fully delegate physical memory allocation and address translation to the hardware. This section compares \sys with other virtual memory designs and related works.}

\textbf{Virtual Memory in Modern Architectures.} 
\sg{Modern virtual memory architectures, such as those employed as part of modern instruction set architectures~\cite{intelx86manual, arm2013, powerpc2015, pa-risc}, have evolved into sophisticated systems.
These architectures have support for features such as large pages, multi-level page tables, hardware-managed TLBs, and variable-size memory segments, but require significant system software support to enable these features and to manage memory. While system software support provides some flexibility to adapt to new ideas, it must communicate with hardware through a rigid \onurii{contract. Such rigid hardware/software communication introduces} costly overheads for many applications (e.g., high overheads with fixed-size \onurii{per-application virtual address spaces,} for applications that only need a small fraction of the space) and \onurii{prevents} the easy adoption of significantly different virtual memory architectures \onurii{or ideas that depend on large changes to the existing virtual memory framework}.
\sys is a completely different framework from existing virtual memory architectures. It supports the functionalities of existing virtual memory architectures, but can do much more by reducing translation overheads, \onurii{inherently and} seamlessly supporting virtual caches, and avoiding unnecessary physical memory allocation.  These benefits come from enabling completely hardware-managed physical memory allocation and address translation, which no modern virtual memory architecture does.}

\onurv{Several memory management frameworks~\cite{vijaykumar2016,etc,vm8,mask,pichai2014,vm9} are designed to minimize the virtual memory overhead in GPUs. Unlike \sys, these works provide optimizations \emph{within the existing virtual memory design}, so their benefits are constrained to the design of conventional virtual memory.}

\textbf{OS Support for Virtual Memory.}
There has been extensive work on how address spaces should be mapped to execution contexts~\cite{lindstrom1995}. Unix-like OSes provide a rigid one-to-one mapping between virtual address spaces and processes~\cite{ritchie1978, mckusick2014}.
SpaceJMP~\cite{vm17} proposes a design in which processes can jump from one virtual address space to another in order to access larger amounts of physical memory.
Single address space OSes \onurvii{rely on system-software-based mechanisms to expose a single global address space to processes}, to facilitate efficient data sharing between processes~\cite{heiser1998, opal, chase1992}. \onurvii{\sys makes use of a similar concept \onurii{as single address space OSes} \nas{with its single globally-visible VBI address space.} However, while existing single address space OS designs expose the single address space to processes, \sys does not do so, and instead has processes operate on CVT-relative virtual addresses. This allows \sys to enjoy the same advantages as single address space OSes (e.g., synonym-/homonym-free VIVT caches), while providing further benefits (e.g., non-fixed addresses for shared libraries, hardware-based memory management). Additionally, \sys naturally supports single address space sharing between the host OS and guest OSes in virtualized environments.}

\textbf{User-Space Memory Management.}
Several OS designs propose user-space techniques to provide an application with more control over memory management\onurii{~\cite{engler1995,avm,vm19,barrelfish,hand1999,fos,ebbrt,sel4, kaashoek1997}}. For example,
\sg{the exokernel OS architecture~\cite{engler1995, kaashoek1997} allows applications to manage their own memory and
provides} memory protection via capabilities, thereby minimizing OS
involvement. Do-It-Yourself Virtual Memory Translation \sg{(DVMT)~\cite{vm19} decouples} memory translation from protection in the OS, and allows applications to handle their virtual-to-physical memory translation. \sg{These solutions (1)~increase application complexity and add non-trivial programmer burden to directly manage hardware resources, and (2)~do not expose \onurii{the rich} runtime information available in the hardware to memory managers. 
In contrast to these works, which continue to rely on software for physical memory management, \sys does not use \emph{any} part of the software stack for physical memory management.  \onurii{By partitioning the duties differently between software and hardware, and, importantly, performing physical memory management in the memory controller,} \sys provides similar flexibility benefits as user-space memory management without introducing additional programmer burden.}

\textbf{Reducing Address Translation Overhead.}
Several studies have characterized the overhead of virtual-to-physical address translation in \sg{modern} systems, which occurs primarily due to growing physical memory sizes, inflexible memory mappings, and virtualization~\cite{vm2, vm40, vm20, intelx86manual, merrifield2016, vm29}. 
Prior works try to ameliorate the address translation issue by: (1)~increasing the TLB reach to address a larger physical address space~\cite{vm6, pham2014, karakostas2015, vm36, vm12, vm37, vm42,vm8}; (2)~using TLB speculation to speed up address translation~\cite{vm38, pham2015tr, papadopoulou2015}; (3)~introducing and optimizing page walk caches to store intermediate page table addresses~\cite{vm1, vm10, vm11, esteve14}; (4)~adding sharing and coherence between caching structures to share relevant address translation updates~\cite{vm33, esteve14, vm6, vm31, vm34, bharadwaj2018, kaxiras2013, latr}; (5)~allocating and using large contiguous regions of memory such as superpages~\cite{vm2, vm3, vm35, vm25, vm8, pham2015}; (6)~improving memory virtualization with large, contiguous memory allocations and better paging structures~\cite{vm25, pham2015, pham2015tr, vm37, vm8, vm35}; and (7)~prioritizing page walk data throughout the memory hierarchy~\cite{mask}. While all \sg{of} these works can mitigate the translation overhead, they build on top of \onurv{the} existing rigid \sg{virtual memory} \onurv{framework} and do not address the underlying \sg{overheads inherent to \onurv{the existing rigid framework} and to software-based memory management. Unlike these works, \sys is a completely new framework for virtual memory, which eliminates several underlying sources of address translation overhead and enables many other benefits (e.g., efficient memory management in virtual machines, easy extensibility to heterogeneous memory systems). \sys can be combined with \onurii{some} of the above proposals to further optimize address translation.}

\section{Conclusion}
\label{sec:conclusion}

\onurii{We introduce the \sysfull (\sys), a new virtual memory framework to address the challenges in adapting conventional virtual memory to increasingly diverse system configurations \onurii{and workloads}.
The key idea \onurv{of} \sys is to delegate memory management to hardware in the memory controller. \nasi{The memory-controller-based memory management in \sys} leads to many benefits not easily attainable in existing virtual memory, such as \nasi{inherently} virtual caches, \nasi{avoiding 2D page walks} in virtual machines, and delayed physical memory allocation.
We \onurv{experimentally} show that \sys (1)~reduces the overheads of address translation by reducing the number of translation requests and associated memory accesses, and
(2)~increases the effectiveness of managing heterogeneous main memory architectures.
We conclude that \sys is a promising new virtual memory framework that can enable several important optimizations and increased design flexibility for virtual memory.
We believe and hope that \sys will open up a new direction and many opportunities for future work \onurv{in novel virtual memory frameworks}.}

\section*{Acknowledgments}

\onurii{We thank the anonymous reviewers of ISCA 2019, MICRO 2019, HPCA 2019, and ISCA 2020 for their valuable comments. We thank our industrial partners, especially Alibaba, Facebook, Google, Huawei, Intel, Microsoft, and VMware, for their generous donations. We thank SAFARI group members for valuable feedback and the stimulating environment.}

{
\bstctlcite{IEEEexample:BSTcontrol} 
\let\OLDthebibliography\thebibliography
  \renewcommand\thebibliography[1]{
    \OLDthebibliography{#1}
    \setlength{\parskip}{0pt}
    \setlength{\itemsep}{0pt}
  }
  \bibliographystyle{IEEEtranS}
  \bibliography{ms}
}

\end{document}